%% file: VHzQ_arXiv_v2.tex
\documentclass[usenatbib]{mnras}

\input {format}

\begin{document}

\title[Infrared properties of Very high-$z$ Quasars]
  {The Near and Mid-infrared photometric properties of known redshift $z\geq5$ Quasars}
\author[Ross \& Cross]
{Nicholas P. Ross\thanks{E-mail: npross@roe.ac.uk} and Nicholas J. G. Cross \\ 
Institute for Astronomy, University of Edinburgh, Royal Observatory, Edinburgh, EH9 3HJ, United Kingdom\\ }

\maketitle
\begin{abstract}
We assemble a catalogue of 488 spectroscopically confirmed very high
($z\geq5.00$) redshift quasars and report their near- ($ZYJHK_{s}/K$)
and mid- (WISE W1234) infrared properties. 97\% of the VH$z$Q sample
is detected in one or more NIR ($ZYJHK/K_{s}$) band, with lack of
coverage rather than lack of depth being the reason for the
non-detections. 389 (80\%) of the very high redshift quasars are
detected at 3.4$\mu$m in the W1 band from the unWISE catalog and all
of the $z\geq7$ quasars are detected in both unWISE W1 and W2. Using
archival WFCAM/UKIRT and VIRCAM/VISTA data we check for photometric
variability that might be expected from super-Eddington accretion. We
find 28 of the quasars have sufficient NIR measurements and
signal-to-noise ratio to look for variability. Weak variability was detected
in multiple bands of SDSS J0959+0227, and very marginally in the
$Y$-band of MMT J0215-0529. Only one quasar, SDSS J0349+0034, shows
significant differences between WFCAM and VISTA magnitudes in one band.
With supermassive black hole accretion likely to be redshift invariant
up to very high-redshift, further monitoring of these sources is
warranted.  All the data, analysis codes and plots used and generated
here can be found at:
\href{https://github.com/d80b2t/VHzQ}{\tt github.com/d80b2t/VHzQ}.
\end{abstract}


\section{Introduction}
Very high redshift quasars (VH$z$Q; defined here to have redshifts
$z\geq5.00$) are excellent probes of the early Universe. This includes
studies of the Epoch of Reionization for hydrogen \citep[see e.g.][for
reviews]{Fan2006review, Mortlock2016}, the formation and build-up of
supermassive black holes \citep[e.g., ][]{Rees1984, WyitheLoeb2003,
Volonteri2010, Agarwal2016, Valiante2018, Latif2018, Wise2019} and
early metal enrichment \citep[see e.g., ][]{Simcoe2012, Chen2017,
Bosman2017}.

Super-critical accretion, where $\dot{M} > \dot{M}_{\rm Edd}$, is a
viable mechanism to explain the high, potentially super-Eddington,
luminosity and rapid growth of supermassive black holes in the early
universe \citep[e.g.,][]{AlexanderNatarajan2014, MadauHaardtDotti2014,
Volonteri2015, Pezzulli2016, Lupi2016, Pezzulli2017, Takeo2018}. Thus,
one could expect VH$z$Qs to potentially vary in luminosity as they go
through phases of super-critical accretion. These signatures of
photometric variability should be looked for, noting the rest-frame
optical emission is redshifted into the observed near-infrared (NIR)
at redshifts $z>5$, and relativistic time dilation also stretches the
photometric variability of the accretion disk as observed.
Fortunately, data are now in place from deep, wide-field NIR
instruments and surveys such as the Wide Field Camera (WFCAM)
instrument on the United Kingdom Infra-Red Telescope (UKIRT) in the
Northern Hemisphere and the VISTA InfraRed CAMera (VIRCAM) on the
Visible and Infrared Survey Telescope for Astronomy (VISTA) in the
Southern Hemisphere, that are necessary for identifying VH$z$Qs.

Quasars are known to be prodigious emitters of infrared emission,
thought to be from the thermal emission of dust grains heated by
continuum emission from the accretion disc
\citep[e.g.,][]{Richards2006b, Leipski2014, Hill2014,
Hickox2017}. Observations in the mid-infrared, e.g. $\sim$3-30$\mu$m
allow discrimination between AGN\footnote{Historically, ``quasars''
and ``Active Galactic Nuclei (AGN)'' have described different
luminosity/classes of objects. In recognition of the fact that both
terms describe accreting supermassive black holes, we use these terms
interchangeably, with a preference for quasar, since we are generally
in the higher-$L$ regime \citep[e.g.][]{Haardt2016book}.}  and passive
galaxies due to the 1.6$\mu$m ``bump'' entering the MIR at
$z\approx0.8-0.9$ \citep[e.g., ][]{Wright1994, Sawicki2002, Lacy2004,
Stern2005, Richards2006b, Timlin2016} as well as between AGN and
star-forming galaxies due to the presence of Polycyclic Aromatic
Hydrocarbon (PAHs) at $\lambda >3\mu$m \citep[e.g., ][]{Yan2007,
Tielens2008}.

\citet{Jiang2006dust} and \citet{Jiang2010} report on the discovery of
a quasar without hot-dust emission in a sample of 21 $z\approx6$
quasars. Such apparently hot-dust-free quasars have no counterparts at
low redshift. Moreover, those authors demonstrate that the hot-dust
abundance in the 21 quasars builds up in tandem with the growth of the
central black hole. But understanding how dust first forms and appears
in the central engine remains an open question \citep{WangR2008,
WangR2011}.

WISE mapped the sky in 4 passbands, in bands centered at wavelengths
of 3.4, 4.6, 12, and 23$\mu$m. The all sky `AllWISE' catalogue
release, contains nearly 750 million detections at
high-significance\footnote{\href{wise2.ipac.caltech.edu/docs/release/allwise/expsup/sec2\_1.html}{wise2.ipac.caltech.edu/docs/release/allwise/expsup/sec2\_1.html}},
of which over 4.5M AGN candidates have been identified with 90\%
reliability \citep{Assef2018}.  \citet{Blain2013} presented WISE
mid-infrared (MIR) detections of 17 (55\%) of the then known 31
quasars at $z > 6$. However, \citet{Blain2013} was compiled with the
WISE `All-Sky' data release, as opposed to the superior AllWISE
catalogues. That sample only examined the 31 known $z>6$ quasars; our
sample has 174 objects with redshift $z \geq 6.00$ (with 117 detected
in WISE). \citet{Banados2016} reports WISE W1, W2, W3 and W4
magnitudes for the Panoramic Survey Telescope and Rapid Response
System 1 \citep[Pan-STARRS1, PS1;][]{Kaiser2002, Kaiser2010}, but with
no further investigation into the reddest WISE waveband for the
VH$z$Qs.

Critically, we now have available to us new W1 and W2 photometry from
the `unWISE Source Catalog' \citep[][]{Schlafly2019}, a WISE-selected
catalogue that is based on significantly deeper imaging and has a more
extensive modeling of crowded regions than the AllWISE release. For
the first time in a catalogue, unWISE takes advantage of the ongoing
mid-IR Near-Earth Object Wide-Field Infrared Survey Explorer
Reactivation mission \citep[NEOWISE-R; ][]{Mainzer2014}, and achieves
depths $\sim$0.7 mag deeper than AllWISE (in W1/2).  This additional
depth is a significant advantage in the detection and study of VH$z$Qs
in the 3-5 micron regime.

Here we present for the first time the combined near-infrared
properties (from UKIRT and VIRCAM) and the new mid-infrared unWISE for
all the spectroscopically known $z\geq5.00$ quasars. Our motivations
are numerous and include: {\it (i)} establishing the first complete
catalogue of $z>5.00$ quasars since the pioneering work from SDSS;
{\it (ii)} utilizing all the WFCAM and VISTA near-infrared photometry
available for the quasars; {\it (iii)} making the first study of near-
and mid-IR variability of the VHzQ population and {\it (iv)}
establishing the photometric properties for upcoming surveys and
telescopes including the Large Synoptic Survey Telescope
(LSST)\footnote{\href{https://www.lsst.org}{lsst.org};}, ESA {\it
Euclid}\footnote{\href{https://sci.esa.int/euclid/}{sci.esa.int/euclid};}
and the {\it James Webb Space Telescope}
(JWST)\footnote{\href{https://www.jwst.nasa.gov/}{jwst.nasa.gov};}$^,$\footnote{\href{https://sci.esa.int/jwst/}{sci.esa.int/jwst};}$^,$\footnote{\href{https://www.asc-csa.gc.ca/eng/satellites/jwst/}{www.asc-csa.gc.ca/eng/satellites/jwst};}$^,$\footnote{\href{https://jwst.stsci.edu/}{jwst.stsci.edu}.}. We
chose redshift $z=5.00$ as our lower redshift limit due to a
combination of garnishing a large sample, adequately spanning physical
properties (e.g. luminosity, age of the Universe) and to highlight the
parts of $L-z$ parameter space where $z>5$ quasars still wait to be
discovered.

This paper is organized as follows. In Section 2, we present the
assembled list of the 488 $z\geq5.00$ VH$z$Qs that we have
compiled. We then give a high-level overview of the photometric
surveys and datasets we use and present the photometry of the
VH$z$Qs. In Section 3 we investigate the variability properties of the
VH$z$Qs, looking for evidence of super-critical accretion. In Section
4 we calculate how many $z>5$ quasars we should expect to find in
current datasets. We conclude in Section 5 and present all the
necessary details to obtain our dataset in the Appendices.

We present all our photometry and magnitudes on the AB system.  For NIR, we use
the \citet{Oke_Gunn1983} and \citet{Fukugita1996} system which has  a zeropoint
of $-48.60$. For WISE, the AB-Vega offsets
\href{http://wise2.ipac.caltech.edu/docs/release/allsky/expsup/sec4_4h.html#conv2ab}{are calculated} 
using the \citet{TokunagaVacca2005} AB system, which has a zeropoint of -48.574.
So to remain on the -48.60 system, and assuming a flux density F$_{\nu} \propto
\nu^{0}$,  one needs to subtract 0.026 mag from the zeropoints:
($\Delta$W1, $\Delta$W2, $\Delta$W3) = (2.673, 3.313, 5.148). 
For W4, we use the calibration of \citet{Brown2014b}, $\Delta$W4= 6.66. 
These magnitudes are {\it not} Galactic extinction corrected. 

We use a flat
$\Lambda$CDM cosmology with $H_{\rm 0} = 67.7$ km s$^{-1}$ Mpc$^{-1}$,
$\Omega_{\rm M} = 0.307$, and $\Omega_{\Lambda} = 0.693$
\citet{Planck2016} to be consistent with \citet{Banados2016} and all
logarithms are to the base 10.

\begin{figure}
  \includegraphics[width=8.6cm, clip,trim=32mm 4mm 32mm 10mm]
  {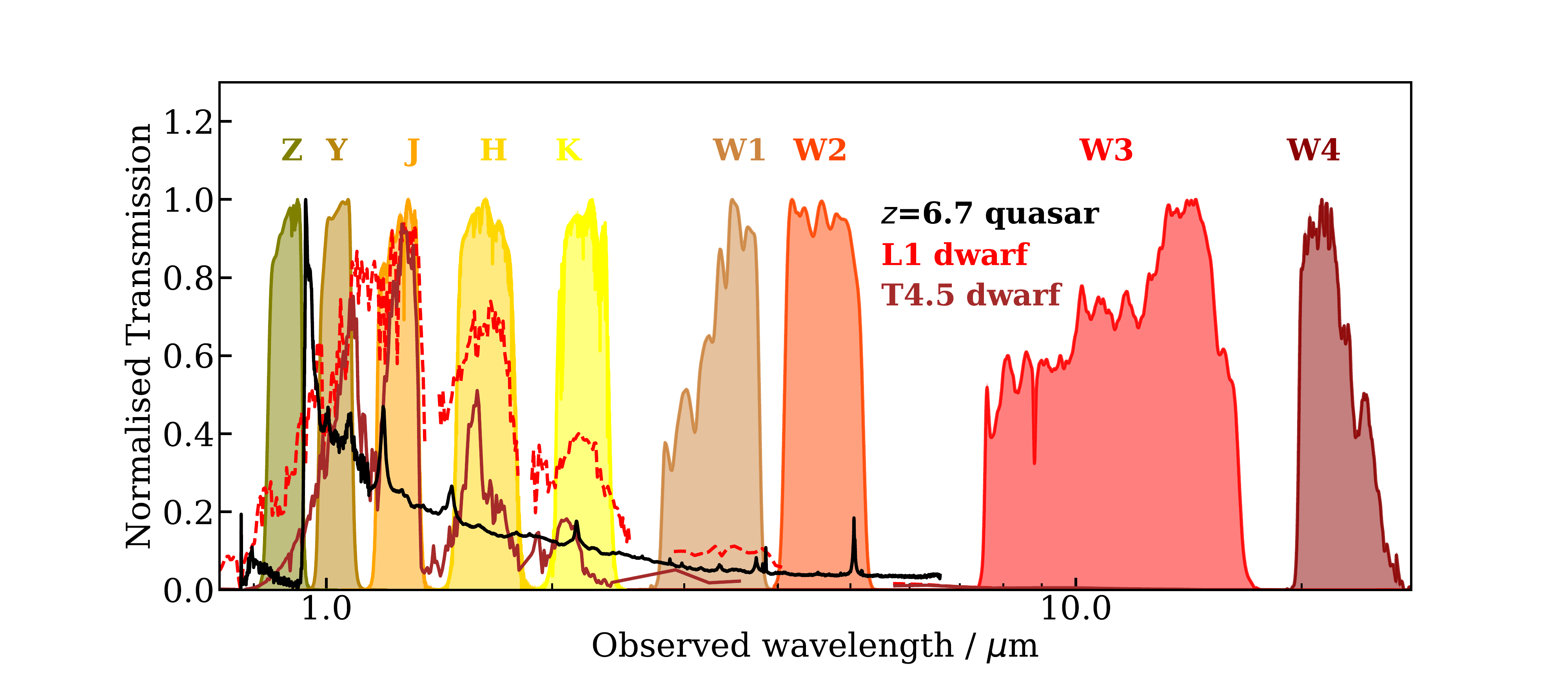}
  \centering
  \vspace{-12pt}
  \caption[]
  {The spectral bands used in this paper. 
    The $ZYJHK$ filter curves are from UKIRT/WFCAM.  
    The $ZYJHK_{s}$ VISTA/VIRCAM filter curves are similar, but not identical to these. 
    The main difference is the $K_s$ band in VISTA that is narrower than the
    WFCAM K, but there are other minor differences. The two sets of filters are
    shown in
    \href{https://github.com/d80b2t/VHzQ/blob/master/SEDs/filters$\_$vs$\_$QSOstars$\_$UKIRT.png}{\tt
    github.com/d80b2t/VHzQ/SEDs/filters$\_$vs$\_$QSOstars$\_$UKIRT.png} and 
    \href{https://github.com/d80b2t/VHzQ/blob/master/SEDs/filters$\_$vs$\_$QSOstars$\_$VISTA.png}{\tt
    github.com/d80b2t/VHzQ/SEDs/filters$\_$vs$\_$QSOstars$\_$VISTA.png}
    respectively.
    The WISE passbands W1-4 are presented, though without the \citet{Brown2014b} W4 recalibration.
    The quasar spectrum is a composite based on \citet{VdB2001} and 
    \citet{Banados2016}. The L and T dwarf spectra are from \citet{Cushing2006}. }
  \label{fig:filters}
\end{figure}

\vspace{-16pt}
\section{Method and Data}
Quasars are generally identified by photometric selection followed by
spectroscopic confirmation. Here, we reverse this method obtaining
first a list of spectroscopic quasars and then obtain photometric
information.

We have compiled a list of 488 quasars with redshifts $z\geq5.00$. We
use all the $z\geq5.00$ quasars that have been discovered,
spectroscopically confirmed and published as of 2018 December 31 (MJD
58483). We then obtain optical, near-infrared and mid-infrared
photometry for the spectral dataset. The near-infrared data comes from
two sources: first, the WFCAM \citep[][]{Casali2007} on the UKIRT,
primarily, but not exclusively, as part of the UKIRT Infrared Deep Sky
Survey \citep[UKIDSS; ][]{Lawrence2007}.  And second, data from the
VIRCAM on the VISTA \citep[][]{Emerson2006, Dalton2006}. The
mid-infrared, $\lambda=3-30\mu$m wavelength data is from the 
Wide-Field Infrared Survey Explorer \citep[WISE;][]{Wright2010,
Cutri2013} mission. For reference, Figure~\ref{fig:filters} displays
the wavelength and normalised transmission of the filters in question.

\subsection{Spectroscopy} 
We compile the list of all known, spectroscopically confirmed
quasars from the literature.  This list was complied from a range of
surveys and papers.  Specifically, we use data from:
\citet{Banados2014, Banados2016, Banados2018a, Banados2018b}, 
\citet{Becker2015}, 
\citet{Calura2014}, 
\citet{Carilli2007, Carilli2010}, 
\citet{Carnall2015}, 
\citet{Cool2006}, 
\citet{Douglas2007}, 
\citet{DeRosa2011}, 
\citet{Fan2000, Fan2001c, Fan2003, Fan2004, Fan2006, Fan2018}, 
\citet{Goto2006}, 
\citet{Ikeda2017}, 
\citet{Jiang2008, Jiang2009, Jiang2015, Jiang2016},   
\citet{Kashikawa2015}, 
\citet{Khorunzhev2017}, 
\citet{Koptelova2017}, 
\citet{Kozlowski2019}, 
\citet{Kim2015, Kim2018},  
\citet{Kurk2007, Kurk2009}, 
\citet{Leipski2014}, 
\citet{Mahabal2005}, 
\citet{Matsuoka2016,  Matsuoka2018a, Matsuoka2018b},   
\citet{Mazzucchelli2017}, 
\citet{Morganson2012}, 
\citet{Mortlock2009, Mortlock2011},
\citet{McGreer2006, McGreer2013},  
\citet{Reed2015, Reed2017}, 
\citet{Saxena2018},
\citet{Schneider2007}, 
\citet{Stern2007},  
\citet{Tang2017}, 
\citet{vanBreugel1999}, 
\citet{Venemans2007, Venemans2012, Venemans2013, Venemans2015a, Venemans2015b, Venemans2016},
\citet{WangF2016, WangF2017, WangF2018a, WangF2018b},
\citet{Willott2007, Willott2009, Willott2010a, Willott2013b, Willott2015}, 
\citet{Wu2015} 
\citet{YangJ2018a, YangJ2018b}  
and 
\citet{Zeimann2011}.

We updated our list of objects between the submitted paper and accepted due
to feedback from other scientists. We discuss these changes in detail in
Appendix~\ref{app:qso_sample}.

Most of these objects are easily identified by their broad Ly$\alpha$
emission line, \nv emission and characteristic shape blueward of
1215\AA\ in the rest-frame. As we shall see, some of the recently
discovered objects are close to the galaxy luminosity function
characteristic luminosity $M^{*}$, and some have relatively weak or
maybe even completely absorbed Ly$\alpha$ \citep[e.g. Figures 7 and 10
in][]{Banados2016}. We leave aside detailed investigation and
discussion into spectral features and line strengths, and take as
given the published spectra and redshift identifications.

The breakdown of how many VH$z$Q each survey reports is given in
Table~\ref{tab:surveys}. The Sloan Digital Sky Survey (SDSS) and the
Pan-STARRS1 (PS1; PSO in Table~\ref{tab:surveys}) survey and alone
identified over half (55.5\%) of the VH$z$Q population. Data from the
Hyper Suprime-Cam (HSC) on the Subaru telescope is responsible for
13.1\% of our dataset (HSC+SHELLQs in Table~\ref{tab:surveys}). The
combination of surveys is also vital for identifying VH$z$Qs. The
UKIDSS Large Area Survey (ULAS) on its own, or in combination with
other surveys is responsible for 6.3\% of the sample (SUV+ULAS)
including the highest-$z$ object. Where more than one survey is used
for the high-redshift identification (e.g. via shorter-band veto and
longer wavelength detection) we follow the discovery paper naming
convention.

\begin{table}
\begin{tabular}{l r r l}
\hline  \hline
Survey              & \# VH$z$Qs & (\%) & Survey reference  \\
\hline  
  ATLAS             &     4    &   ( 0.82)    &  \citet{Shanks2015} \\
  CFHQS            &   20    &   ( 4.10)    &  \citet{Willott2007} \\
  DELS$^{a}$       &   16    &   ( 3.28)    &  \citet{Dey2018} \\
  ELAIS              &     1    &   ( 0.20)    &  \citet{Vaisanen2000} \\
  FIRST              &     1    &   ( 0.20)    &  \citet{Becker1995} \\
  HSC                 &    9    &   ( 1.84)    & \citet{Miyazaki2018} \\
  IMS                 &     5    &   ( 1.02)     &  \citet{Kim2015} \\
  MMT               &   12    &   ( 2.46)     &  \citet{McGreer2013} \\
  NDWFS           &     1    &   ( 0.20)    &  \citet{JD1999} \\
  PSO                 &   100   &   (20.49)   &   \citet{Kaiser2002, Kaiser2010}
  \\
  RD                   &     1   &   ( 0.20)    &  \citet{Mahabal2005} \\
  SDSS                &  171  &    (35.04)    & \citet{EDR} \\
 SDWISE$^{b}$    &   27    &  ( 5.53)    &   \citet{WangF2016} \\
  SHELLQs         &    55    &   (11.27)  &  \citet{Matsuoka2016}     \\  
  SUV$^{c}$       &   21     &    (4.30)  & \citet{YangJ2017} \\
  TGSS				&	1		&	(0.20)	& \citet{Saxena2018} \\
  TN				& 	1		&	(0.20)	& ?? \citet{vanBreugel1999} \\
  UHS               &    1      &  ( 0.20)     &  \citet{WangF2017} \\
  ULAS               &   10   &   ( 2.05)     & \citet{Lawrence2007} \\
  VDES$^{d}$       &   17  &    ( 3.48)     &  \citet{Reed2017} \\
  VHS                 &     1  &     ( 0.20)    & \citet{WangF2018b} \\
  VIK                 &     10    &  ( 2.05)    &  \citet{Edge2013} \\
  VIMOS           &    1      &  ( 0.20)     &   \citet{LeFevre2003} \\
  XMM				& 	1	&	 (0.20)		& \citet{Khorunzhev2017} \\
\hline  \hline
\end{tabular}
\caption{The source and number of the VH$z$Q, with the key survey reference also given. 
  Recent survey name and acronyms include: 
  $^{a}$DESI Legacy Imaging Survey; 
  $^{b}$SDWISE = SDSS+WISE; 
  $^{c}$SUV  = SDSS+ULAS/VHS; 
  $^{d}$VDES = VHS/VIKING+DES; 
}
      \label{tab:surveys}
\end{table}

\begin{figure}
  \includegraphics[width=8.0cm, clip, trim=10mm 0mm 0mm 0mm]
   {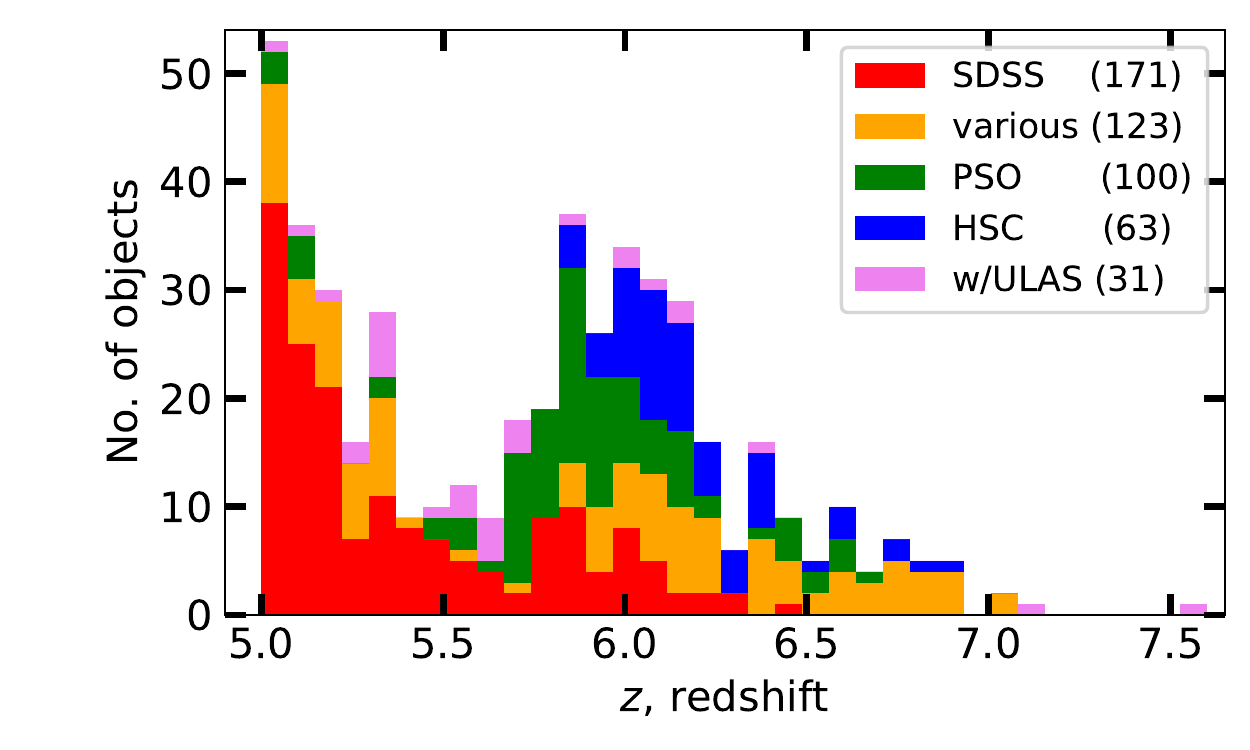}
  \centering
  \vspace{-12pt}
  \caption[]
  {The redshift distribution $N(z)$ of the VH$z$Q sample. 
    The bins are $\delta z=0.075$ in width and have the data 
    stacked on top of each other.}
  \label{fig:Nofz}
\end{figure}

The redshifts for the VH$z$Qs generally come from the measurement of
broad UV/optical emission lines. There are far infra-red emission
lines e.g. \cii 158 $\mu$m available for several objects, but at the
level of our current analysis broadline redshifts are sufficient.

\begin{table}
\centering
\begin{tabular}{l r  r}
\hline \hline
$z \geq$  & Age / Myr & No. of objects \\
\hline 
5.00         & 1180          &  488   \\
5.70         & 1000          &  273  \\
6.00         &   937          &  174   \\
6.19         &   900          &  87   \\
6.50         &    845         &  40   \\
6.78         &    800         &  14   \\
7.00         &    767         &   4   \\
7.50         &    700         &   1   \\
\hline \hline
\end{tabular}
\caption{The number of objects at or above a given redshift. 
The age of the Universe in Megayears is also given. }
      \label{tab:ages}
\end{table}

The number of objects at or above various redshifts, along with the 
corresponding age of the Universe is given in Table~\ref{tab:ages}. 

The $N(z)$ redshift histogram is given for the sample in Figure~\ref{fig:Nofz}. 
We split the contribution up by survey. For clarity we show the individual 
surveys of SDSS, PS1, HSC, the ULAS detection, and tally the remaining 
surveys together (``various'').

\begin{figure}
  \includegraphics[width=8.0cm, clip, trim=0mm 0mm 0mm 10mm]
 {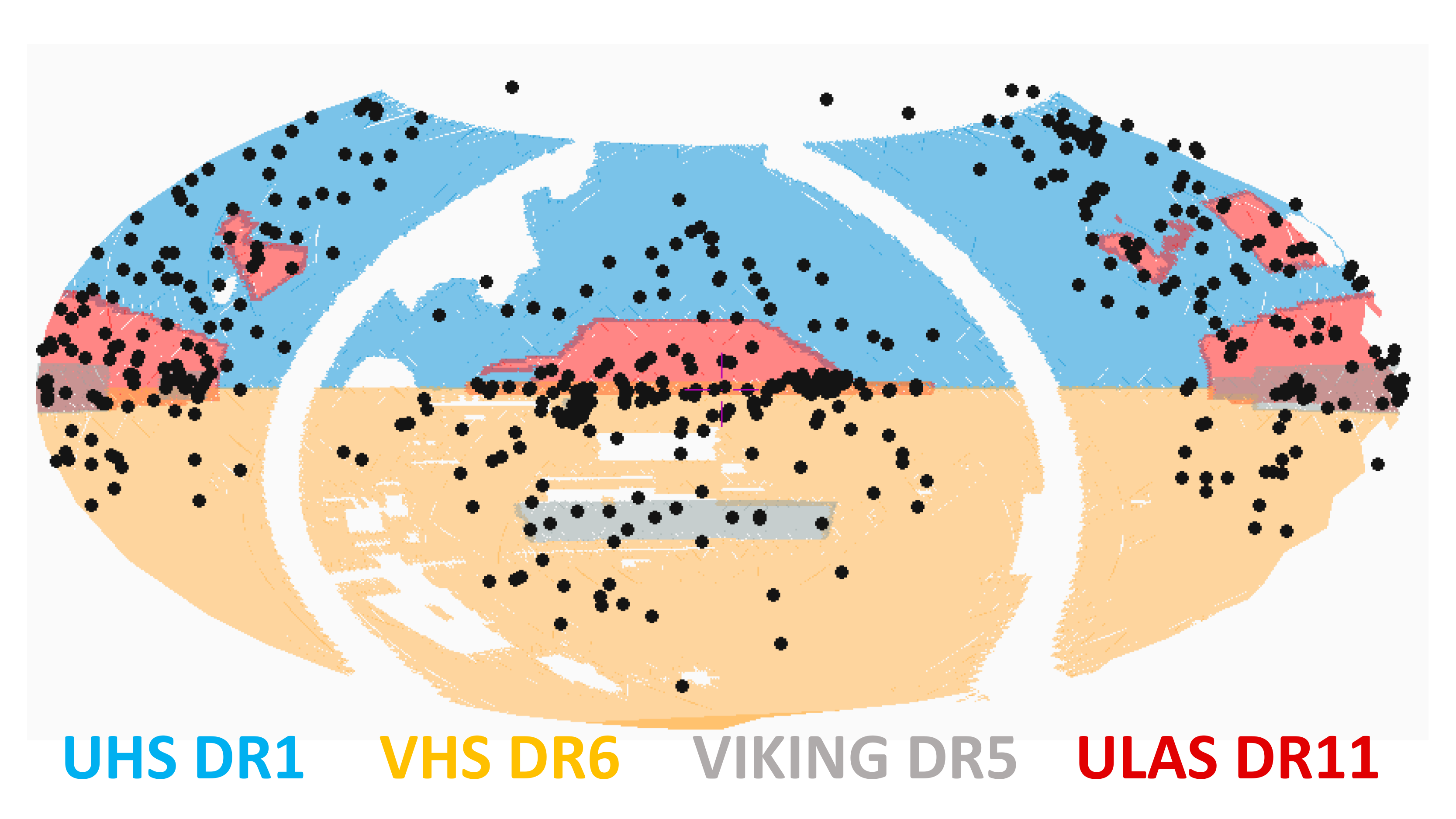}
  \centering
  \caption[]
  {The coverage maps for VHS (DR6; orange), UHS (DR1; olive), VIKING
(DR5; blue) and ULAS (DR11; red), the most recent public releases of
the 4 main surveys. The VH$z$Qs are given by black dots.}
  \label{fig:coverage}
\end{figure}

\subsection{Near-infrared photometry}~\label{sec:NIR_data} 
The near-infrared data in this paper comes from the Wide Field
Astronomy Unit's \href{https://www.roe.ac.uk/ifa/wfau/}{(WFAU)}
Science Archives for UKIRT-WFCAM, the WFCAM Science Archive
\citep[WSA; ][]{Hambly2008} and VISTA-VIRCAM, the VISTA Science
Archive \citep[VSA; ][]{Cross2012}. These archives were developed for
the VISTA Data Flow System \citep[VDFS][]{VDFS}.

We access both the WSA and the VSA and include all non-proprietary
WFCAM data, which covers all public surveys and PI projects from
Semester 05A (2005-05-01) to 2017-Jan-01, and all non-proprietary VISTA data, which
covers all public surveys and PI projects from science verification on
2009-Oct-15 to 2017-Jan-01.

Here we are not just querying the WSA or VSA data tables. We are
taking a list of objects (positions) are performing matched aperture
(``forced'') photometry on the NIR imaging data. As such, we generate
a set of tables that are different in subtle ways to the regular
``Detection'' tables.  The two most important tables for our needs are
the {\tt [w/v]serv1000MapRemeasurement} and {\tt
[w/v]serv1000MapRemeasAver}.

We produce and provide a two new databases with all the necessary
quantities and measurements to fully reproduce our tables, figures and
results herein. Moreover, these databases report considerably more
information than we report here. Full documentation can be found at
the \href{http://wsa.roe.ac.uk/www/wsa_browser.html}{WSA Schema
Browser} and the
\href{http://horus.roe.ac.uk/vsa/www/vsa_browser.html}{VSA Schema
Browser}.

Figure~\ref{fig:coverage} shows the areal coverage of the ULAS, UHS,
VHS and VIKING. UHS DR1 is 12,600 deg$^{2}$ ($J$-band); ULAS DR1 3,700
deg$^{2}$ VHS DR6 16,000 deg$^{2}$ and VIKING DR5 is 1,300
deg$^{2}$. The overlap between UHS DR1 and VHS DR6 is 28 deg. These
four surveys together cover 33,000 deg$^{2}$. VHS, ULAS and UHS do not
observe in the $Z$-band.

  \subsubsection{Averaging matched photometry}
  The data was processed using a matched-aperture photometry method
  where flux is measured at the spectroscopic position of the quasar,
  without necessarily knowing if there is a formal detection in the NIR
  photometry beforehand. The matched-aperture pipeline is discussed in
  \citet{Cross2013} and with fuller details to appear in a forthcoming
  paper (Cross et al., 2020, in prep).

  We query the WSA and VSA performing matched-aperture photometry at
  the positions of our 488 VH$z$Qs. This database is world-readable and
  we give the full recipe and relevant SQL queries for accessing both
  databases in Appendix~\ref{sec:SQL_WSA} as well as online. 

  The photometry in a single epoch image often has a low
  signal-to-noise ratio. The advantage of matched aperture photometry on
  quasars is that co-adding is relatively simple if each epoch is taken
  in the same aperture and the aperture photometry has been corrected to
  total. Indeed, the standard aperture corrections work well for point
  sources. Coadding using the matched-aperture photometry can give
  better results for the photometry, where the individual epochs are
  taken from multiple projects with different field-centres and
  orientations and point-spread functions since the individual epoch
  scattered light, pixel distortion and aperture corrections can be
  applied with the correct weighting.

  We average the aperture corrected calibrated fluxes (e.g. {\bf
  aperJky3}), and then convert to magnitudes. 

  \begin{equation}
    \bar{F} = \frac{\sum_i^N (w_i\,F_i)}{\sum_i^N w_i}  
    \label{eq:avg}
  \end{equation}
  where $F_i$ is the $i^{th}$ epoch measurement of a parameter to be
  averaged such as the aperture corrected calibrated flux in a $1\arcsec$ aperture
  ({\bf aperJky3}) and $\bar{F}$ is the weighted mean average of this parameter.
  The weight for each epoch $w_i=1/(\sigma_{F})^2$ if the epoch is included and 
  $w_i=0$ if an epoch is excluded for quality control purposes. The weights of each epoch in each averaged catalogue are tabulated in the {\tt [w\v]serv1000MapAverageWeights}.

  We calculate a set of averaged catalogues, for each pointing and filter, based
  on the requirements in \verb+RequiredMapAverages+, in these cases over time
  spans of 7, 14, 30, 91, 183 days, 365 days, 730 days, over 10 epochs and
  over all epochs. The averaging process starts at the first epoch and works onwards 
  from there. Again, we present these measurements in the new SQL tables. 

  We detect 367 unique quasars in the WFCAM WSA database, 237 quasars
  are detected in the VISTA VSA database with 131 objects in common with
  both WFCAM and VISTA data.  We give the necessary SQL queries syntax
  at \href{https://github.com/d80b2t/VHzQ/blob/master/data/WSA_VSA/SAMPLE_SQL_QUERIES}{\tt
    d80b2t/VHzQ}.

\subsection{MIR data}
The MIR data for this study comes from the Wide-field Infrared Survey Explorer
(WISE) mission, and we utlize data from the WISE cryogenic and the Near-Earth
Object WISE \citep[NEOWISE; ][]{Mainzer2011} post-cryogenic and NEOWISE
Reactivation Mission \citep[NEOWISE-R][]{Mainzer2014} survey phases.

We use data from the beginning of the WISE mission \citep[2010 January;
][]{Wright2010} through the fifth-year of NEOWISE-R operations \citep[2018
December;]{Mainzer2011}. There are several  major
\href{https://irsa.ipac.caltech.edu/Missions/wise.html}{data releases} and  
catalogues based on the WISE mission. Here we use two: the WISE 
\href{http://wise2.ipac.caltech.edu/docs/release/allwise/}{AllWISE Data release}
and the recently released ``unWISE Catalog'' \citep{Schlafly2019}. The AllWISE 
program combines the W1 and W2 Single-exposure data from all the WISE survey
phases  (4-Band Cryo, 3-Band Cryo and Post-Cryo; 2010-01-07 thru 2011-02-01)
survey  phases, and the W3 and W4 from the 4-Band Cryo phase. The  unWISE
effort\footnote{\href{http://unwise.me}{http://unwise.me}} is the unblurred 
coadds of the WISE imaging using the AllWISE and NEOWISE-R stacked data 
\citep{Lang2014, Meisner2018a, Meisner2018b}.

For the two shorter WISE bands, $\lambda_{\rm eff}$3.37$\mu$m W1 and 
$\lambda_{\rm eff}$4.62$\mu$m W2 we generally report the deeper, unblurred
unWISE  coadd data.  For the two longer WISE bands, $\lambda_{\rm
eff}$12.1$\mu$m  and $\lambda_{\rm eff}$22.8$\mu$m we use the AllWISE Data
Release. Most objects in the AllWISE Source Catalog are unresolved, so the best 
photometric measurements to use are the deep detection profile-fit photometry
measures,   {\tt w$x$mpro},  {\tt w$x$sigmpro} and  {\tt w$x$snr}. The unWISE
Catalog  absolute photometric calibration derives from the photometric
calibration  of the unWISE coadds \citep{Meisner2017a}, which is tied to the 
original WISE zero points through aperture fluxes in a 27.5'' radius.

Previous works \citep[e.g., ][]{Krawczyk2013, Ross2015, Bilicki2016} found  that
cross-matches performed with a radius of 2-3'' between the user catalogue and 
WISE was a good compromise between completeness and contamination  \citep[see
e.g. Figure 4 of ][]{Krawczyk2013}. We thus use a cross-match radius of 2.75''. 
When querying the AllWISE catalogues, ``Cone Search Radius'' in the AllWISE 
table search was set to 2.75'' for the Spatial Constraints. The ``One to One
Match''  was ``not'' checked; although possible, we consider it highly unlikely 
there would be more than one MIR source contributing to the flux of a single 
UV/optically bright rest-frame quasar. As a check, we utilize data from the 
{\it Spitzer} Space Telescope in IRAC Channel 2 4.5$\mu$m,  IRAC Channel4
8$\mu$m and MIPS 24$\mu$m,  and number counts and luminosity density
measurements  from \citet[][see their Figures 8.11, 8.12 and 8.13]{Driver2016}.
The redshift distribution of the sampled 8--24 $\mu$m foreground objects is 
narrow enough and their $K$-corrections advantageous enough, that the objects 
counts sharply turnover for $AB(W3, W4) \gtrsim 20$mag, and the normalized 
differential counts --- especially in e.g. WISE W4 --- have most of their  
extragalactic background light energy emerging at AB$\sim$18--20 mag. As a 
consequence, the relative number or probability of contaminating objects at 
AB$>20$mag inside our apertures should be small.  

Knowing we have secure detections in the near-infrared bands, and wanting  to
boost the number of WISE W3/W4 detections, we allow ourselves to be less 
conservative in querying the AllWISE catalogues and also query the AllWISE 
Reject Table. However, with the exception of one object (SHELLQs  J1208-0200), 
the AllWISE Reject Table does not contain any further W3/W4 detection information.

All fluxes in the unWISE catalog are reported there are in ``Vega
nanoMaggies'', with the Vega magnitude of a source is given by
\begin{equation}
m_{\rm Vega} = 22.5 - 2.5\log(f),
\end{equation} 
where $f$ is the source flux. The absolute calibration for unWISE is
ultimately inherited from AllWISE through the calibration of
\citet{Meisner2017a}. This inheritance depends on details of the PSF
normalization at large radii, which is uncertain. Subtracting 4
millimag from the unWISE W1, and 32 millimag from unWISE W2 fluxes
improves the agreement between unWISE and AllWISE fluxes.

Thus to convert unWISE Vega magnitudes onto the AB system, we have:     
\begin{eqnarray*}
        {\rm W1}_{\rm AB, unWISE}  & = &   22.5 - 2.5 \log(f_{\rm W1}) - 0.004 + 2.673  \\
       {\rm W2}_{\rm AB, unWISE}  &  = &  22.5 - 2.5 \log(f_{\rm W2}) - 0.032 +  3.313 \\
\end{eqnarray*}
using the \citet{TokunagaVacca2005} AB system, 
and is slightly different to the transformations presented in 
\citet{Schlafly2019}. 

For the our MIR variability investigations, we use the 
\href{http://wise2.ipac.caltech.edu/docs/release/neowise/}{NEOWISE-R 2019 Data
Release}.  NEOWISE 2019 makes available the 3.4 and 4.6 μm (W1 and W2)
single-exposure images  and extracted source information that was acquired up
until  2018 December 13 (MJD 58465) including the fifth year of survey
operations of  NEOWISE. These fifth year NEOWISE data products are concatenated 
with those from the first four years into a single archive from 2013 December 13 (MJD 56639).
The WISE scan pattern leads to coverage of the full-sky approximately once every
six months (a``sky pass''), but the satellite was placed in hibernation in 2011 
February and then reactivated in 2013 October. Hence, our light curves have a
cadence  of 6 months with a 32 month sampling gap.

\begin{landscape}
 \input{output_table_top5pc}

\end{landscape}

\section{Results}
Having collated the sample of 488 VH$z$Qs, and obtained their near-
and mid-infrared photometry we report here the various photometric
properties of the quasars.
Table~\ref{tab:output_table} represents the culmination of this
effort, {\it is the main
\href{https://github.com/d80b2t/VHzQ/blob/master/data/VHzQs_ZYJHK_WISE.dat}{data
product} of this paper} and we now exam the assembly of its contents
in more detail. 

First, we will concentrate on detection rate in the infrared, go on to
report on the colour-redshift and colour-colour properties of our sample. 
We also investigate the photometric variability. 

\subsection{Detection Rates in the NIR}
Table~\ref{tab:nir_detection} gives the detection rates for the 
VH$z$Qs in the NIR $YJHK/K_{s}$-bands. 
The first thing to note is that the coverage of the NIR surveys 
for example from the UKIDSS LAS and VISTA VHS, does
not overlap the full area for where the VH$z$Qs are detected. 

\begin{table}
  \centering
  \begin{tabular}{l r l}
    \hline  \hline
    Selection   & number detected (\%) \\
    \hline  
    Any band ($ZYJHK/K_{s}$)    &  473  (96.9) \\
    $Z$-band                    &  81  (16.6) \\
    $Y$-band                    &  289  (59.2) \\
    $J$-band                    &  471  (96.5) \\
    $H$-band                    &  277  (56.8) \\
    $K$ or $Ks$-band            &  343  (70.3) \\
    \hline  \hline
  \end{tabular}
  \caption{Detection rate of VH$z$Qs in the near-infrared. 
For the 15 objects that have no NIR detections, 3 have 
been observed but are not in our queried time range, 6 have 
not been observed yet and 6 objects are too far north to be visible 
by UKIRT. }
  \label{tab:nir_detection}
\end{table}

There are 15 objects that have no NIR detections. 3 of these
(PSOJ053.9605-15.7956, PSO J056.7168-16.4769 and DELSJ0411-0907) have
been observed (by VHS) but are out of our queried time range (for
which the data is publicly available). 6 objects are at declination
$\delta<0$ deg and have not been observed (or at least the data is not
in the VSA archive yet). 6 objects are at a declination $\delta \geq
+60$ deg are too far north for UKIRT and cannot be observed.

We note that although we are often working close to the magnitude
limits of the surveys, Eddington bias does not a significant effect on
our measurements, since we are not discovering new sources, but
measuring properties of already known quasars. Indeed in the NIR
WFCAM/VISTA data, using forced photometry reduces the Eddington bias
to negligible levels since a flux will be measured regardless of
whether the source is just bright enough (or too faint) to be detected
above the background noise.

  \subsubsection{Comparing WFCAM and VISTA}
  There are 131 quasars observed in the overlapping area between WFCAM
  and VISTA. We used the {\tt VegaToAB} values to put these objects on
  the same AB system, and for each object compared the two
  measurements. First, we calculated the weighted average (calibrated
  flux) in each filter of both and calculated the ratio and difference
  between each measurement and the average.  Then for each filter we
  calculated the weighted average of the differences (in mag) for each
  instrument to see if there were significant offsets. The results are
  given in Table~\ref{tab:WFCAM_vs_VISTA}.  The only filter with a
  significant offset is the $Y$-band. All of the VISTA averages are
  negative and all of the WFCAM ones are positive, except for J where the
  difference is negligable. The $K_s$ versus $K$ band comparison may be affected
  the different shapes of the filters, with $K$ being significantly wider than $K_s$, able to detect light at
  longer wavelengths.
  \begin{table}
    \centering
    \begin{tabular}{l r r}
      \hline  \hline
      abs(VIRCAM & \multirow{2}{*}{millimags} &  no. of  \\
      -  WFCAM)      &                                        &  objects \\
      \hline
      $Z$                 &  13.1 	& 3 \\
      $Y$                 &  59.3 	& 63 \\
      $J$                  &    0.9 	& 125 \\
      $H$                 &  14.9     &  108 \\
      $K_{\rm s}$/$K$ &  1.1     & 114 \\
      \hline  \hline
    \end{tabular}
    \caption{Comparing the magnitudes in different WFCAM/UKIRT and 
      VIRCAM/VISTA near-infrared bands.}
    \label{tab:WFCAM_vs_VISTA}
  \end{table}

  We have also checked for quasars with large differences in magnitude
  between the WSA and VSA. Objects were selected where the average flux
  was $>0.$ and $>5$ average flux error. In order to account for large
  errors in either the WFCAM or VISTA photometry, objects with
  $\delta$mag$ < 2 \times$ $\sigma_{\delta\rm mag}$ in either WSA or VSA
  were removed. This selects one quasar with large, $\delta$mag
  $>0.2$mag differences between the WSA and VSA. This is SDSSJ0349+0034
  which has a WSA $K$-band magnitude of 21.03$\pm$0.35 mag AB (observed at
  MJD=54062.4), while for the VSA $K_s$-band this is 20.20$\pm$0.14 mag AB
  (observed 3 times between MJD=55537 and 55542), meaning that the WFCAM and
  VISTA observations are $\sim4$ years apart.

\begin{table}
  \centering
  \begin{tabular}{l r r r r}
    \hline \hline
    \multirow{ 2}{*}{$z \geq$} & \multicolumn{2}{c}{unWISE} &  \multicolumn{2}{c}{AllWISE}   \\ 
                                             & \multicolumn{1}{c}{W1} & \multicolumn{1}{c}{W2} & \multicolumn{1}{c}{W3} & \multicolumn{1}{c}{W4}  \\ 
    \hline
    5.00   &   389 ( 79.7\%)   &   334 ( 68.4\%)  &    59 ( 12.1\%)  &    10 ( 
    2.0\%) \\
    5.70   &   194 ( 71.1\%)   &   157 ( 57.5\%)  &    16 (  5.9\%)  &     6 ( 
    2.2\%) \\
    6.00   &   115 ( 66.1\%)   &    96 ( 55.2\%)  &     9 (  5.2\%)  &     2 ( 
    1.1\%) \\
    6.19   &    65 ( 74.7\%)   &    48 ( 55.2\%)  &     3 (  3.4\%)  &     2 ( 
    2.3\%) \\
    6.50   &    35 ( 87.5\%)   &    26 ( 65.0\%)  &     1 (  2.5\%)  &     2 ( 
    5.0\%) \\
    6.78   &    12 ( 85.7\%)   &     9 ( 64.3\%)  &     0 (  0.0\%)  &     0 (  0.0\%) \\ 
    7.00   &     4 (100.0\%)   &     4 (100.0\%)  &     0 (  0.0\%)  &     0 (  0.0\%) \\ 
    7.50   &     1 (100.0\%)   &     1 (100.0\%)  &     0 (  0.0\%)  &     0 (  0.0\%) \\ 
    \hline \hline
  \end{tabular}
  \caption{Detection rates in the mid-infrared bands from WISE as a function of 
    redshift. A signal-to-noise ratio cut of {\tt w$x$snr}$>3.0$ has been used
    for each band.}
  \label{tab:mir_detection}
\end{table}

\subsection{Detection Rates in the MIR}
Unlike the NIR coverage, the WISE satellite and mission performed an all-sky
survey,  so the location of every VH$z$Q in our dataset is covered. However, the
depth  of the WISE survey depends heavily on sky location, with locations near
the  Ecliptic Poles having the highest number of exposures.

\begin{figure}
  \includegraphics[width=8.6cm, clip,trim=6mm 6mm 0mm 6mm]
 {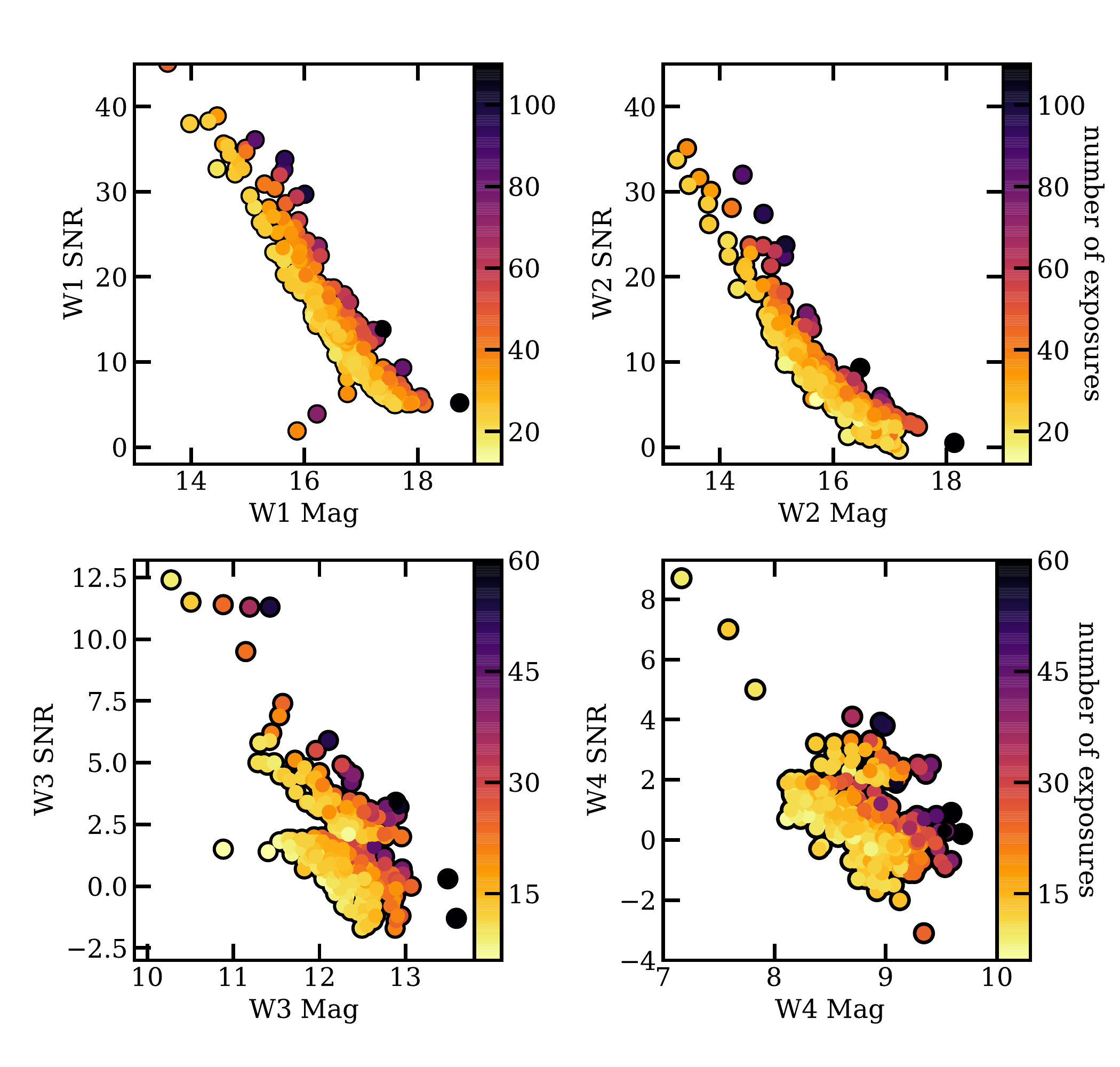}
  \centering
  \vspace{-14pt}
  \caption[]{WISE W1/2/3/4 {\tt w$x$mpro} Vega magnitude against signal-to-noise
  ratio, colour coded by {\tt w$x$cov} the mean coverage depth, in each
  corresponding band.}
  \label{fig:WISEmag_vs_coverage}
\end{figure}

Before reporting on the detection rates, we investigate this effect
using the AllWISE Source Catalog. Figure~\ref{fig:WISEmag_vs_coverage} 
shows the WISE AllWISE magnitude versus signal-to-noise ratio, colour coded
by {\tt w$x$cov} the mean coverage depth, in each corresponding
band. The {\tt w$123$cov} values are the mean pixel coverage in W1/2/3
from the W1/2/3 Atlas Tile Coverage Map within an aperture of circular
area with a radius of 8.25'' centered on the position of this
source. For {\tt w$4$cov} this radius is 16.5'' (the AllWISE Source
Catalog
\href{http://wise2.ipac.caltech.edu/docs/release/allwise/expsup/sec2_1a.html#w2cov}{Column
Descriptions} has further details). The {\tt w$x$cov} value takes into
account e.g., individual pixels in the measurement area that may be
masked or otherwise unusable (reducing the effective pixel count and
thus the mean coverage value) as well as pixels that are affected by
distortions across the across the focal plane in single-exposure
images (where this distortion is corrected when coadding to generate
the Atlas Images). 

In the two shorter bands W1/2 we see the clear and expected
trend for brighter objects to have larger SNR, and also for the higher
signal-to-noise ratio for objects with more exposures at a given
magnitude. The behaviour for the W3/4 bands is different, with two
populations clearly evident in W3 and although a bit more mixed, also
in W4. With the suggested split at SNR$>2$, and no obvious
R.A./Declination dependence seen, this behaviour is explained by the
fact that there are non-detections in W3/4 for objects (with high W1/2
SNR) that are reported in the AllWISE catalogue.

Table~\ref{tab:mir_detection} gives the detection rates for the
VH$z$Qs in the MIR WISE W1-4 bands. The unWISE depths are impressive
with nearly 80\% of all the VH$z$Qs being detected in unWISE W1. 12
out of 14 (86\%) in unWISE W1 (9 our of 14; 64\% in unWISE W2) of the
$z\geq6.78$ quasars are detected and moreover {\it all of the $z\geq7$
quasars are detected in both unWISE W1 and W2}. This bodes very well
for future, small mirror infrared missions \citep[e.g.][]{NEOCam_WP}.

\begin{figure}
    \centering
    \includegraphics[width=8.6cm,  clip, trim=6mm 4mm 6mm 10mm]
    {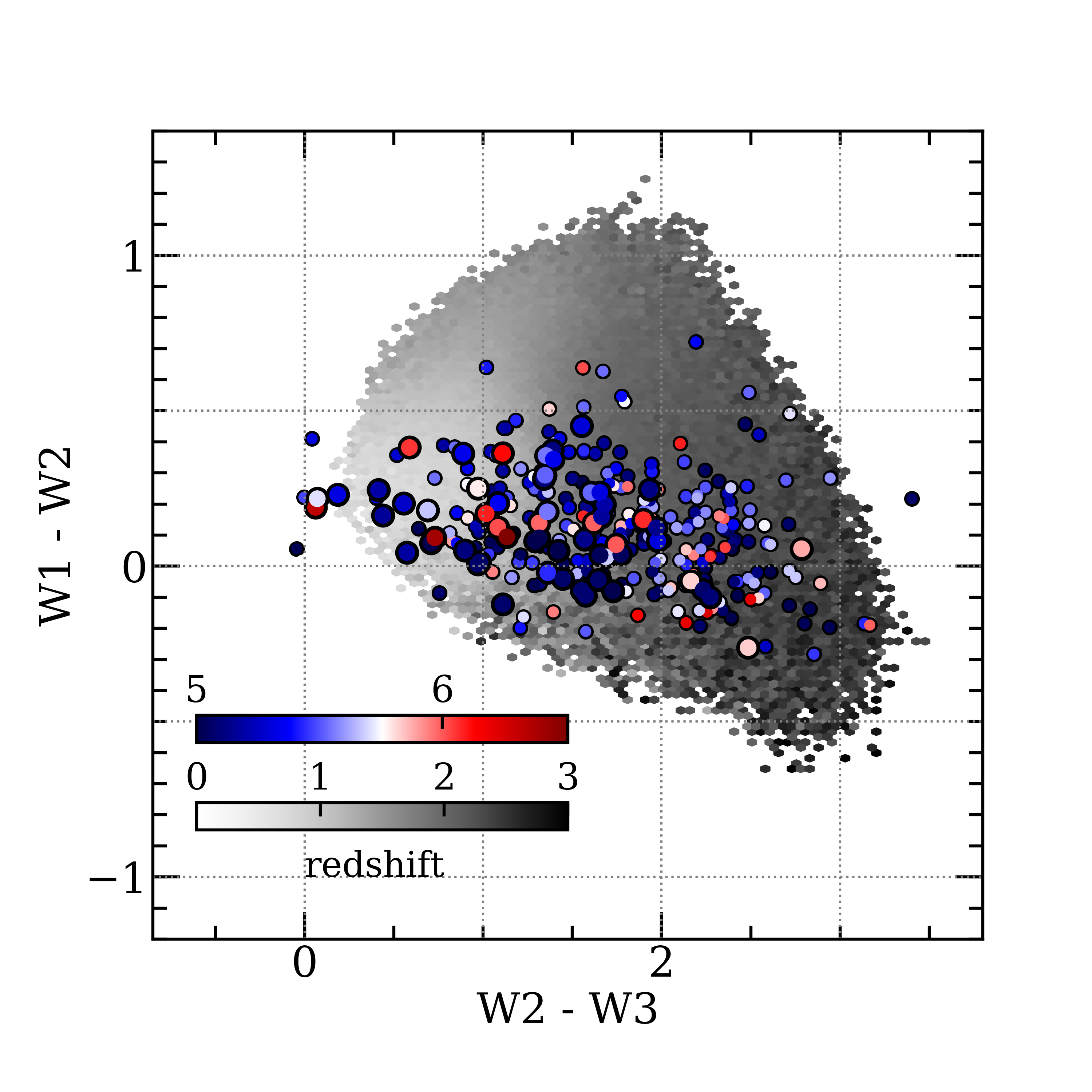}
  \vspace{-14pt}
    \caption{The (W2-W3) vs. (W1-W2) colour – colour diagram showing the WISE colours 
      for the 330 VH$z$Q that have reported w3mpro values (blue to red coloured
      points).
      400,000 quasars at redshift $z\lesssim3.0$ from the SDSS DR14 quasar catalogue 
      \citep{Paris2018} are also shown
      (grey colour-scale). }
    \label{fig:W1W2W3}
\end{figure}

\citet{Secrest2015} present an all-sky sample of $\approx$1.35 million
AGN meeting a two-color, (W1-W2) vs. (W2-W3), photometric selection as
applied to sources from AllWISE. This colour-selection is motivated by
\citet{Mateos2012}. We investigate if this catalogue, which is tuned
for, and noted to be highly complete at, lower, $z\lesssim2$
redshifts, contain the VH$z$Qs. We find only 5 VH$z$Qs from our sample
match the Secrest catalogue, but given a W3 detection is required,
this is not surprising. Further, in addition to the relative lack of
sensitivity in W3, the mid-IR colors are moving out of the
\citet{Mateos2012} color criterion at high redshift. This was shown
using an AGN SED where AGNs above a redshift of about $\sim$2 fall out
of the colour criterion, see e.g. Figure 5 of \citet[][]{Mateos2012}
and Figure 9 of \citet{Secrest2015}.
\citet{Assef2018} released two large catalogues of AGN
candidates identified across 30,000 deg$^2$ of extragalactic sky from
the WISE AllWISE Data Release. The ``R90'' catalogue, contains 4.5M
AGN candidates at 90\% reliability ($\approx$150 AGN candidates
per deg$^2$) while the ``C75'' catalogue consists of 20.9M AGN
candidates at 75\% completeness ($\approx$700 AGN candidates per
deg$^2$). Crossmatching our catalogue of 488 VH$z$Qs with these
catalogues, produces 49 matches with the R90 sample and 129 matches
with the C75 sample. Both catalogues unsurprisingly match to the
ultraluminous quasar SDSS J0100+2802 \citep{Wu2015} while the C75, but
not the R90 catalogue matches to the first redshift $z>7$ quasar ULAS
J1120+0641 \citep{Mortlock2011}. Neither the R90 or C75 catalogue matches to the
current highest-redshift object J1342+0928 \citep{Banados2018a}.

\begin{figure*}
    \centering
     \includegraphics[width=0.24\textwidth]{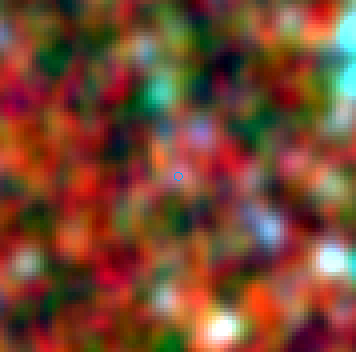}
   \includegraphics[width=0.24\textwidth]{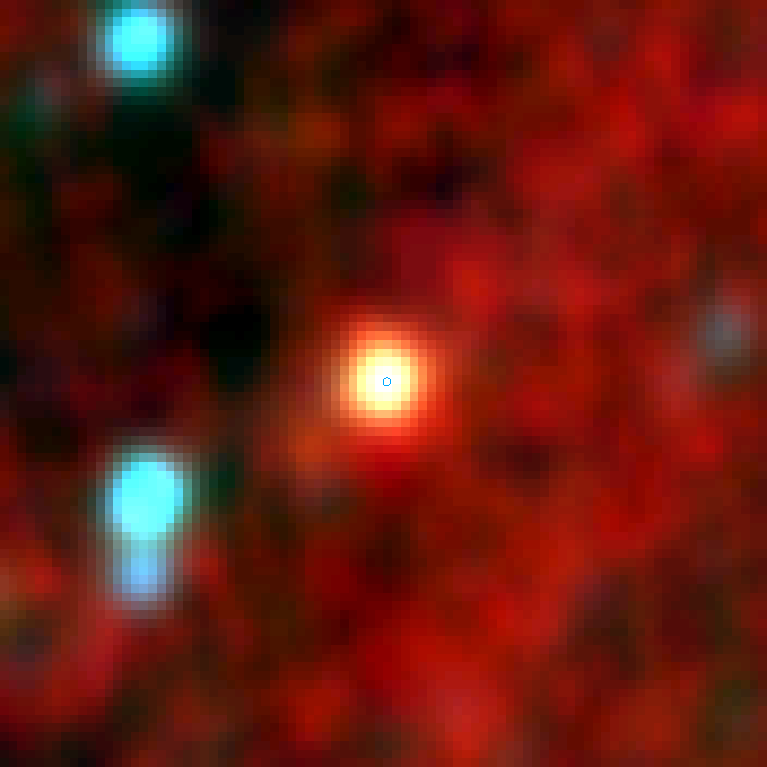} 
    \includegraphics[width=0.24\textwidth]{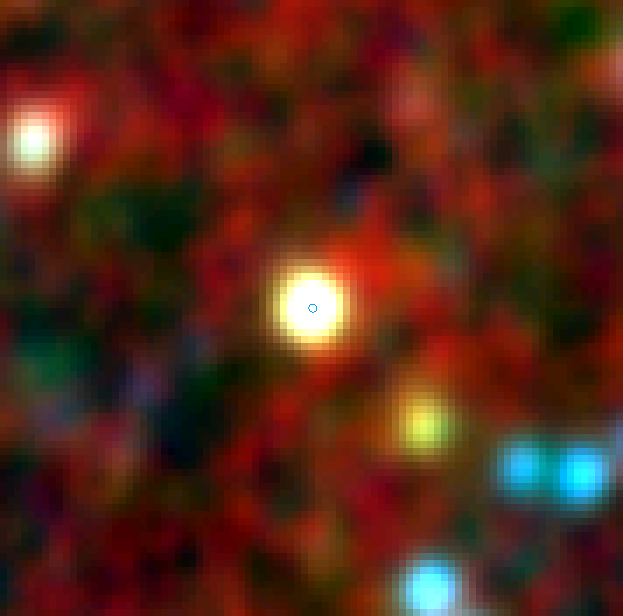}
   \includegraphics[width=0.24\textwidth]{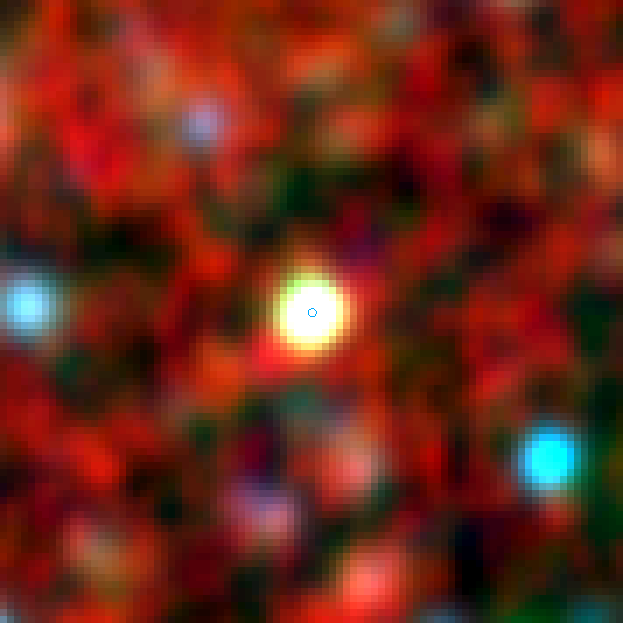}
    \caption{VH$z$Qs with interesting MIR properties. 
      The thumbnails are the RGB colour outputs using the AllWISE W1W2W3 bands from the 
      IRSA WISE Image Service, with a scale of 120'' on the side. 
      {\it (a)} SDSS J0927+2001 is the bluest in (W1-W2) colour quasar.
      {\it (b)} UHS J0439+1634 was discovered by \citet{Fan2019} and is a strongly lensed quasar at $z=6.51$. 
      {\it (c)} SDWISE J0306+1853 has {\tt w3mpro} = $15.66\pm0.1$ (AB) and {\tt w4mpro } = $14.25\pm0.16$ (AB). 
      {\it (d)} SDSS J0100+2802 is the \citet{Wu2015} object.}
    \label{fig:VHzQ_W4s}
\end{figure*}
\subsection{MIR Colours}
Due to the depth all-sky coverage of the WISE (and NEOWISE-(R))
mission, several investigations have looked at how WISE detects AGN
\citep[e.g][]{Stern2012, Assef2012, Secrest2015, LaMassa2017,
Assef2018, Glikman2018, Hviding2018, LaMassa2019}

Figure~\ref{fig:W1W2W3} shows the (W1-W2) vs. (W2-W3) colour-colour
diagram for the 330 VH$z$Qs that have reported {\tt w3mpro} values
(blue to red coloured points), though we note that only 51 (18\%) of
these objects have {\tt w3snr} $>$3.0. Approximately 400,000 quasars
at redshift $z\lesssim3.0$ from the SDSS Quasar Catalog using the
Fourteenth data release \citep[DR14Q; ][]{Paris2018} are also shown
(grey colour-scale).

The VH$z$Q with the bluest colour in (W1-W2) is SDSS~J0927+2001, with
(W1-W2) = --0.283 (i.e. 0.328 in Vega), and the WISE W1W2W3 color image
is shown in Fig.~\ref{fig:VHzQ_W4s} {\it(a)}. The reddest object in
(W1-W2) is PSO~J135.3860+16.2518 with (W1-W2) = 0.721 (1.33
Vega).

The set of VH$z$Q detected in WISE W3 and W4 contains 59 objects that
are detected in the broad W3 filter and 10 objects that are formally
detected in W4 ({\tt w4snr}$\geq$3.0). Three of the W3/4 objects are
presented in Figure~\ref{fig:VHzQ_W4s}. UHS~J0439+1634
(Fig.~\ref{fig:VHzQ_W4s} {\it (b)}) with W4=7.165$\pm$0.12 was
discovered by \citet{Fan2019} and is a strongly lensed quasar at
$z=$6.51. This high luminosity is mostly not intrinsic, but is boosted
by an intervening redshift $z\sim$0.7 galaxy. Other high-redshift
quasars that are bright W3/4 objects e.g. SDWISE~J0306+1853
(Fig.~\ref{fig:VHzQ_W4s} {\it(c)}) may well also so be lensed
\citep[e.g.,][]{Glikman2018lens, Fan2019, Pacucci2019}, but
high-resolution follow-up is needed to confirm. SDSS~J0100+2802
(redshift $z=6.33$; Fig.~\ref{fig:VHzQ_W4s} {\it(d)}) does not have a
formal W4 detection ({\tt w4snr}=2.4), but as reported by
\citet{Wu2015}, has a detection in the 2MASS $J$, $H$ and $K_{\rm s}$
bands and is also the bluest object in both (W1-W3) and (W2-W3).

\begin{figure}
  \centering
  \includegraphics[width=8.5cm]
   {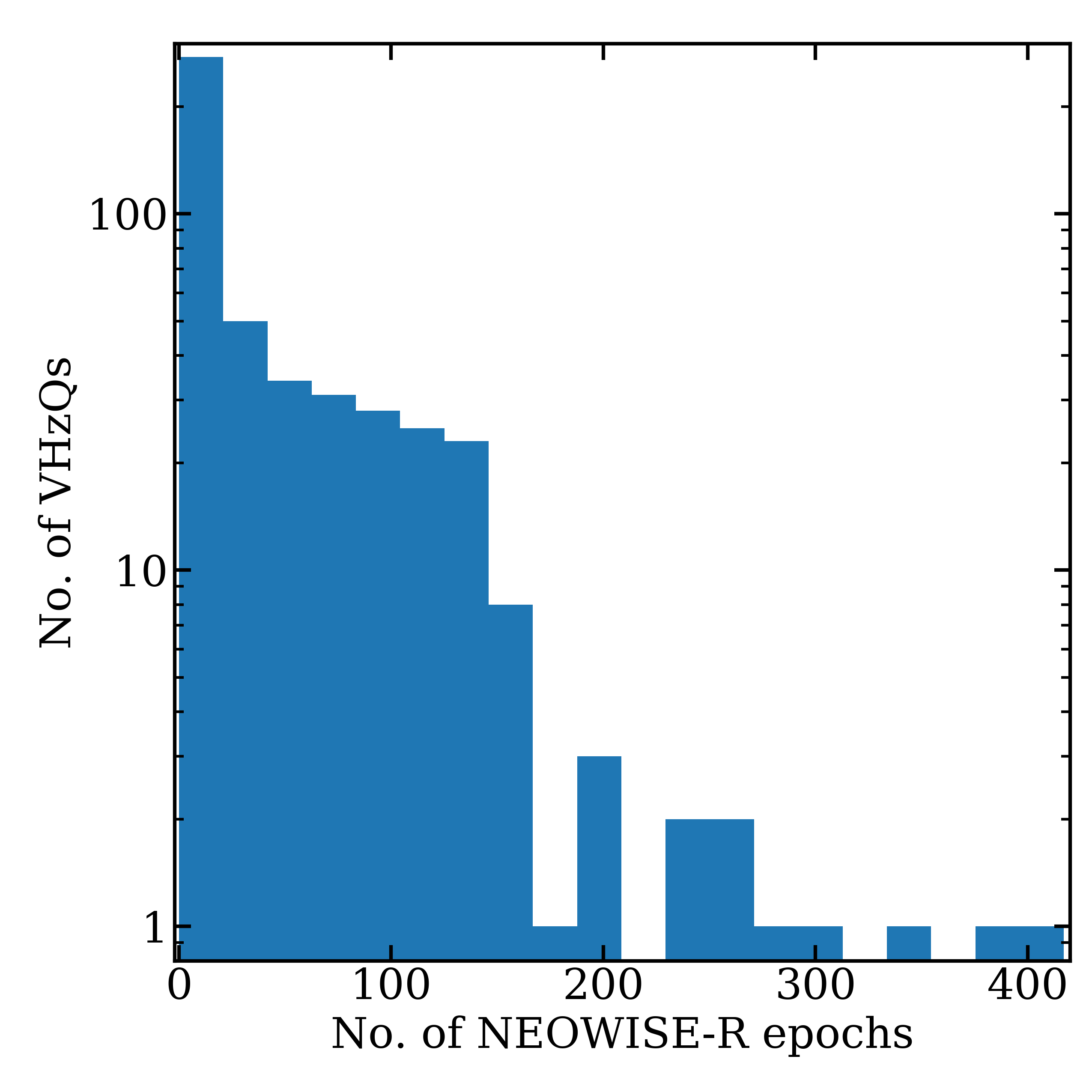}
  \vspace{-16pt}
  \caption[]
  {Histogram showing the total number of exposures for each detected VH$z$Q
    from the full combination of AllWISE, NEOWISE and NEOWISE-R.} 
  \label{fig:MIR_LC_epochs}
\end{figure}
\subsection{Variability}
VH$z$Qs, if accreting at, or above the Eddington limit, might well
have changing mass accretion rates, i.e., $\ddot{m}_{\rm accr}$. A
consequence of this would be that these quasars exhibit signs of
variability, most likely showing up in their UV/optical rest-frame
emission. We look for evidence of this variability signature in the
NIR and MIR light-curves of the VH$z$Qs.

Quasars are known to have dramatically changing Balmer lines,
especially H$\beta$ \citep[e.g.,][]{LaMassa2015, Ruan2016, Runnoe2016,
Macleod2016, Gezari2017, Runco2016, YangQ2018, Assef2018, Stern2018,
Ross2018, MacLeod2019, Graham2019b}. As a guide, we note that H$\alpha$
is redshifted to 3.94$\mu$m (i.e. W1) at $z=5.00$ and 5.57$\mu$m,
which is between the W2 and W3-bands at $z=7.50$.  H$\beta$ is
redshifted to 2.92$\mu$m, which is the blue edge of W1, at $z=5.00$
and 4.13$\mu$m at $z=7.50$. Less well understood is the temporal
behaviour of the metal lines, in particular \civ and \mgii.  \civ
enters the $Y$-band at redshift $z$=5.32 and exits at $z$=5.99, and
enters the $J$-band at redshift $z=6.55$ and exits at $z$=7.57. \mgii
enters the $H$-band at redshift $z=4.33$ and exits at $z=5.37$ and
enters the $K$-band at redshift $z=6.25$ and exits at $7.50$.

\begin{figure}
  \centering
  \includegraphics[width=8.5cm]{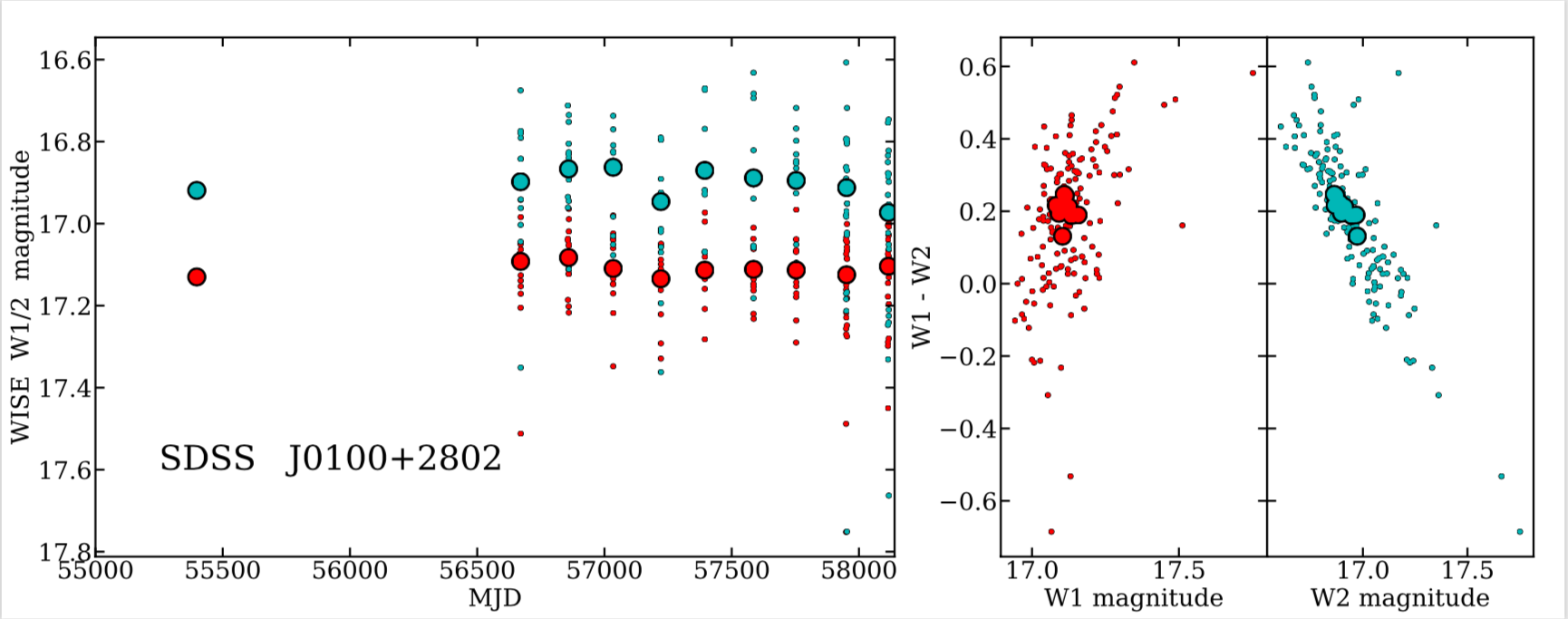}
 \includegraphics[width=8.5cm]{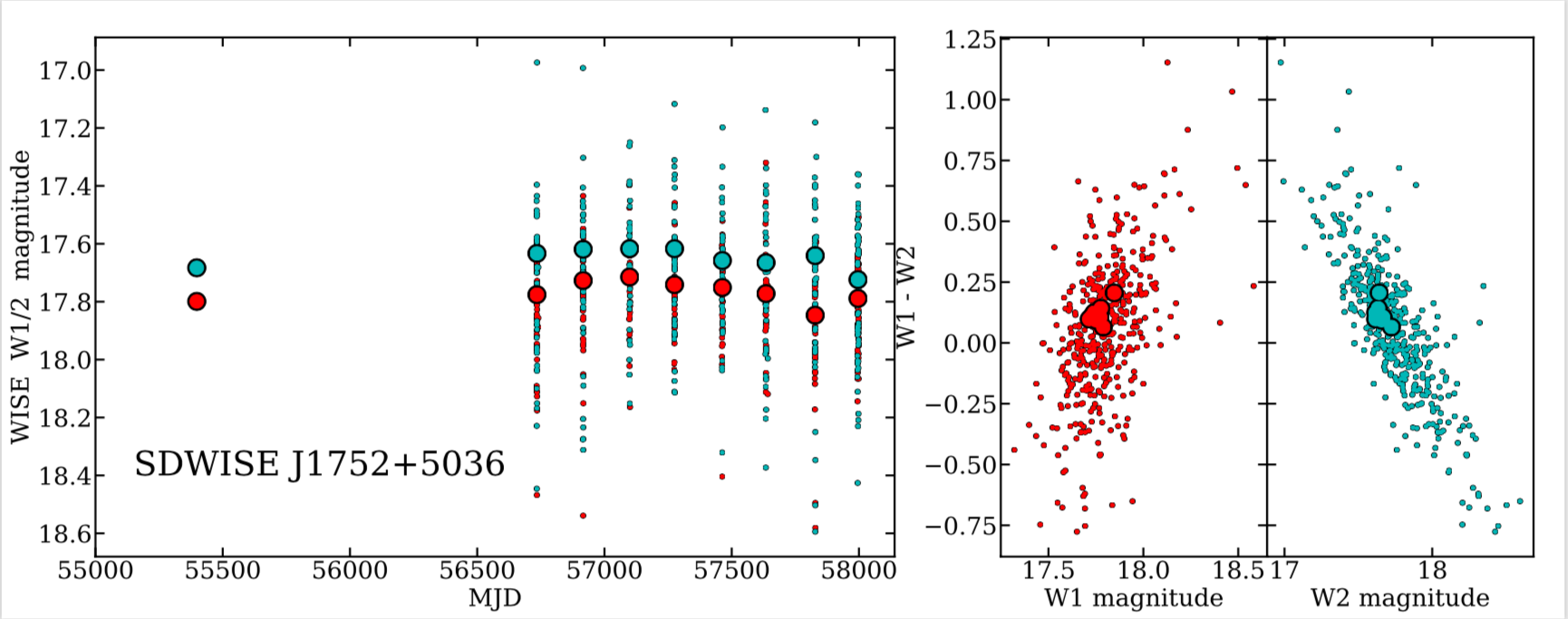}
 \includegraphics[width=8.5cm]{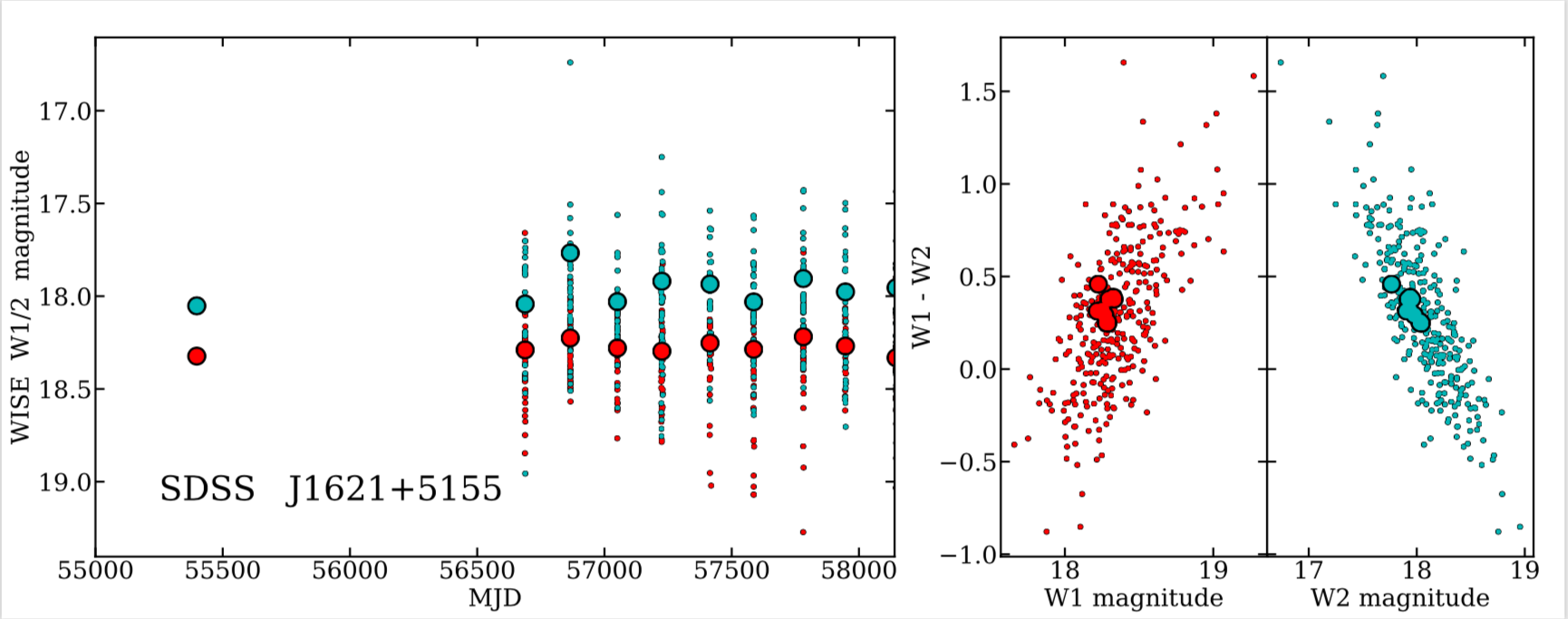}
  \caption[]
  {Here we show the MIR NEOWISE-R for SDSS J0100+2802 (top), SDWISE   J1752+5036 (middle), and  
    SDSS J1621+5155 (bottom). Red points are the W1 band; cyan points the W2 band.
  For the light curves, we show weighted means for every 6 month NEOWISE-R scan.} 
  \label{fig:MIR_LC_3egs}
\end{figure}
One key point to note is that the NEOWISE-AllWISE match radius 
is 3.00'' and the AllWISE (and 2MASS) source information included 
in the NEOWISE Source Database 
\href{http://wise2.ipac.caltech.edu/docs/release/neowise/expsup/sec2_1f.html}{are
associations not identifications}. 

Figure~\ref{fig:MIR_LC_epochs} gives the number of NEOWISE-R epochs and detections for each VH$z$Q. 
There are 74 quasars with over 100 NEOWISE-R epochs, nine with over 200 epochs
and four with over 300.
Figure~\ref{fig:MIR_LC_3egs} presents three examples of the brighter MIR objects with well-sampled 
MIR lightcurves and associated colour changes; 
SDSS J0100+2802 (Fig.~\ref{fig:MIR_LC_3egs} top), 
SDWISE  J1752+5036    (middle), 
SDSS J1621+5155 (bottom). 
For the light curves, we show weighted means for every 6 month NEOWISE-R scan.
Due to the WISE scanning pattern, objects close to the North Ecliptic Pole 
are well sampled, and indeed this is the area for further investigation 
time-domain investigation with upcoming large infrared space telescopes
\citep{Jansen2018, WhiteC2019AAS}. 

Using the extended datasets described in Section~\ref{sec:NIR_data},
we select objects with at least 8 measurements observed with intervals
of at least 30 days, in at least one filter. The the average
calibrated flux over all epochs must be $>0$ (i.e. {\tt aperJky3}$>0$)
with the signal-to-noise ratio is required to be $\geq8$ (i.e. {\tt
aperJky3/aperJky3Err}$\geq$8). With these criteria, there are 18 WFCAM 
and 12 VISTA objects with 2 objects (SDSSJ0221-0342B $\&$ VIK J1148+0056) in
common to both, and this sample of 28 is our starting
point for variability investigations.

\begin{figure}
  \begin{subfigure}{}
    \centering
    \includegraphics[width=8.5cm]{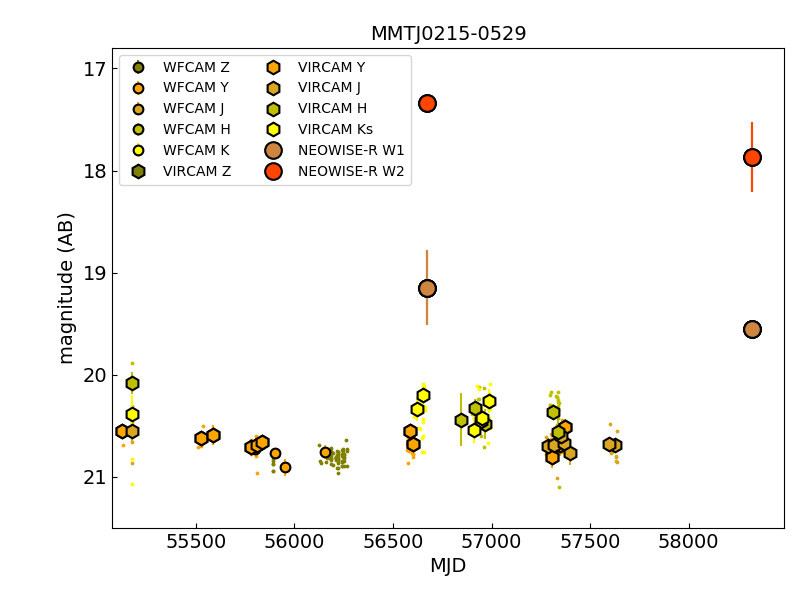}
    \caption{The infrared light curve for MMT J0215-0529 with data from 
      WFCAM (smaller solid circles), VIRCAM (hexagons) and NEOWISE-R (larger solid circles).
MMT  J0215-0529 was identified by \citet{McGreer2018}, 
}
    \label{fig:MMTJ0215-0529}
  \end{subfigure} 
  \begin{subfigure}{}
    \centering
    \includegraphics[width=8.5cm]{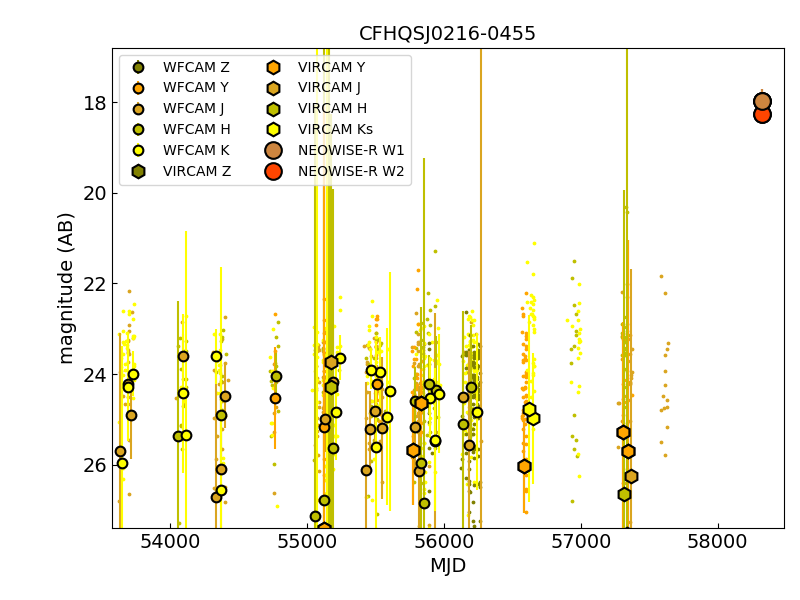}
    \caption{The same as Fig.~\ref{fig:MMTJ0215-0529} except for CFHQS J0216-0455.}
    \label{fig:CFHQSJ0216-0455}
  \end{subfigure}
\end{figure}

\begin{figure}
  \begin{subfigure}{}\quad
    \centering
    \includegraphics[width=8.5cm]{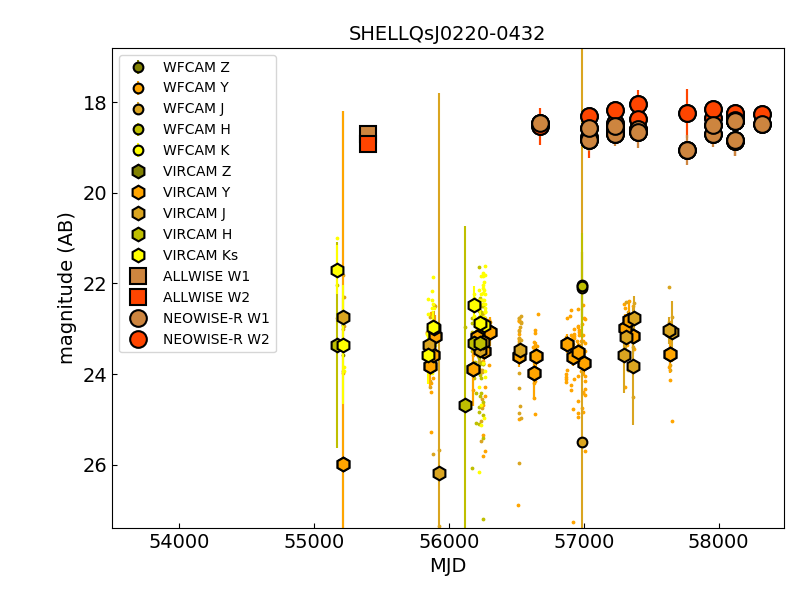}
    \caption{The same as Fig.~\ref{fig:MMTJ0215-0529} except for SHELLQs J0220-0432.}
    \label{fig:SHELLQsJ0220-0432}
  \end{subfigure}
  \begin{subfigure}{}\quad
    \centering
    \includegraphics[width=8.5cm]{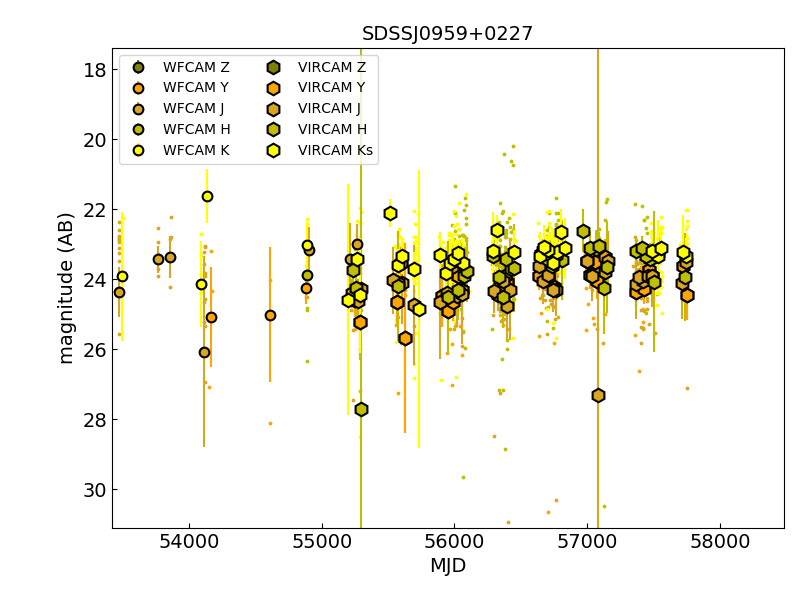}
        \caption{The near infrared light curve for SDSS J0959+0227. Note, 
      SDSS J0959+0227 is too faint in the MIR to be detected by AllWISE or unWISE.}
    \label{fig:SDSSJ0959+0227}
  \end{subfigure}
  \medskip
\end{figure}

The clipped median and median absolute deviation is then calculated
\begin{equation}
      var = 1.48 \times  {\rm m.a.d.} / \bar{\epsilon}
\end{equation}
where $var$ is the index of variation, m.a.d. is the median absolute
deviation, and $\bar{\epsilon}$ is the mean of the error in each point
in the light curve, divided by the total number of points. We apply the criteria 
$var>=3.$ to light-curves from the original measurements and also to light-curves 
where measurements have been averaged over different time-scales to improve 
the signal-to-noise ratio. We found that two quasars showed signs of
variability:
MMT J0215-0529 (see Fig.~\ref{fig:MMTJ0215-0529}) in the $Y$-band with an
average time-scale of 30 days and amplitude of 0.3 mag, and SDSS J0959+0227 
(see Fig.~\ref{fig:SDSSJ0959+0227}) in the $Y$ and $H$-bands, with timescales 
of 1 year or 6 months for $Y$ only. The $H$-band amplitude was 0.7 mag and  the
$Y$-band amplitude was 1.2 mag if averaged over 6 months, or 0.9 mag if averaged over 1 year.

We present four objects that are particularly well-sampled in the NIR. 
These are: 
MMT J0215-0529       ($z=5.13$; Figure~\ref{fig:MMTJ0215-0529}), 
CFHQS J0216-0455    ($z=6.01$; Figure~\ref{fig:CFHQSJ0216-0455}), 
SHELLQs J0220-0432 ($z=5.90$; Figure~\ref{fig:SHELLQsJ0220-0432}) 
and 
SDSS J0959+0227 ($z=5.07$; Figure~\ref{fig:SDSSJ0959+0227}). 
We incorporate the MIR NEOWISE-R light curve data where available. 

We note that sometimes the average flux is negative and so a default
magnitude is reported.  This is apparent for the average $H$ and
$K_{s}$ flux for CFHQS J0216-0455 at MJD=56900.  Interestingly this is
evident in both the $H$ and $K_{s}$ filters, but with this quasar not
much brighter than the detection limit, one must be very careful with
any (over) interpretation.

Even with well sampled data across the 3000-4000 day observed
time-scales that the WFCAM and VISTA surveys span, the $(1+z)$
time-dilation dramatically affects the sampled rest-frame timescales
sampled, which are $\sim$300-700 days. Indeed, when sharp changes in a
accretion rate are expected on the system's dynamical timescale of
several kiloyears \citep[e.g.,][]{Regan2019}, then seeing any
variability signature is not expected. However, noting that {\it (i)} accretion onto 
supermassive black holes should be generally time invariant, and {\it (ii)} 
that in lower redshift quasars, which also have massive $>$$10^{8}$ M$_{\odot}$ black
holes, dramatic changes in both continuum and line emission {\it is
seen} on much shorter timescales, continued monitoring of these
objects is definitely warranted.

We finally focus on SDSS J0959+0227 which is presented in
\citet{WangF2016} but is first reported in \citet{Civano2011} as
CID-2220 and a high-redshift, $z>3$ AGN, in the Chandra-COSMOS
field. The spectrum from this object is presented in \citet{Ikeda2012}
and shows SDSS J0959+0227 having a narrow Ly$\alpha$ line, and would
likely be a Lyman-$\alpha$ emitting galaxy had it not been an X-ray
source, The X-ray luminosity is $\approx$3$\times10^{44}$ erg s$^{-1}$
in the 2-10 keV rest-frame, so it is an AGN. This object is clearly
not a regular broadline ``Type 1'' AGN. Noting that the redshift of
$z=5.07$ for SDSS J0959+0227, the $Y$, $J$, $H$, $Ks/K$-bands
correspond to rest-wavelengths of $\sim$1690, 2055, 2690,
3515/3625\AA\ i.e. the rest-frame UV/very blue, so, so the regular
blue QSO continuum could be emerging. Thus, with the slightest hint of
variability, this could potentially be a high-$z$ AGN transitioning
from a narrow-line ``Type 2'' object to a broadline Type 1 quasar.

\section{Conclusions}\label{sec:conclusions}
In this study, we have, for the first time, compiled the list of all
$z>5$ spectroscopically confirmed quasars. We have assembled the NIR
($Z, Y, J, H, K/K_{s}$) and MIR (WISE W1/2/3/4) photometry for these
objects and given their detection rates. We find that:

\begin{itemize}
  \item SDSS and Pan-STARRS1 together identified over half of the VH$z$Q sample;
  \item There remains a quasar ``redshift desert'' at $z\approx5.3-5.7$,  though
  efforts are being made to address this \citep[e.g.,][]{YangJ2018a};
  \item 96.9\% of the VH$z$Q sample is detected in one or more NIR
    ($ZYJHK/K_{s}$) band;
  \item The 15 objects that are not detected in the NIR are due to
    lack of coverage rather than lack of depth;
  \item 389 (79.7\%) VH$z$Qs are detected by WISE, e.g. in the deeper
    unWISE W1 catalog.
  \item All of the $z\geq7$ quasars are detected in both unWISE W1 and
    W2.
  \item 28 of the quasars had enough NIR measurements and sufficient
    NIR measurements and signal-to-noise ratio to look for variability. Weak
    variability was detected in multiple bands of SDSS J0959+0227, and
    very marginally in the Y-band of MMT J0215-0529. Only one quasar, SDSS
    J0349+0034 had significant differences between their WFCAM and VISTA
    magnitudes in one band, indicating variability.
 \end{itemize}

The science reach of $z>5$ quasars will continue to be important well
into the next decade \citep{Becker2019_DecadalWP, Fan2019_DecadalWP,
Wang2019_DecadalWP} and will provide key insights into direct collapse
black holes, hydrogen reionization and the physics of accretion in the
first $\lesssim$700 million years of the Universe.

\subsection*{Author Contributions}   
N.P.R. initiated the project, compiled the list of $z>5.00$ quasars,
wrote most of the analysis code, developed the plotting scripts, and
developed and wrote the initial and subsequent drafts of the
manuscript.

N.J.G.C. supplied the critical near-infrared expertise and database
for which the bulk of the project relies. N.J.G.C. also contributed
directly to the writing of the manuscript.

\subsection*{Availability of Data and computer analysis codes} 
All materials, databases, data tables and code are fully available at: 
\href{https://github.com/d80b2t/VHzQ}{\tt https://github.com/d80b2t/VHzQ}

\section*{Acknowledgements}
NPR acknowledges support from the STFC and the Ernest Rutherford Fellowship scheme. 

We thank:
\begin{list}{$\circ$}{}  
  \item Mike Read at the ROE WFAU for help with the WFCAM Science Archive (WSA) and the VISTA Science Archive (VSA). 
  \item Aaron Meisner and Eddie Schlafly for facilitating early access to the unWISE Catalog.
  \item Krisztina Perger for key discussions about missing objects after the pre-print appeared. 
  \item Tim Brooke at the \href{https://irsasupport.ipac.caltech.edu/index.php}{IRSA Help Desk}; 
  \item Bernie Shiao at STScI for help with the Pan-STARRS1 DR1 CasJobs interface. 
  \item Michael Cushing for supplying the Late Type stellar spectra and Beth Biller for useful discussion. 
  \item Nathan Secrest for useful discussions on the mid-IR AGN catalogues. 
\end{list}

This paper heavily used \href{http://www.star.bris.ac.uk/~mbt/topcat/}{TOPCAT} (v4.4)
\citep[][]{Taylor2005, Taylor2011}.
This research made use of \href{http://www.astropy.org}{\tt Astropy}, 
a community-developed core Python package for Astronomy 
\citep{AstropyCollaboration2013, AstropyCollaboration2018}.

The VISTA Data Flow System pipeline processing and science archive were used for the WFCAM and VISTA 
near infrared data are described in Irwin et al (2004), Hambly et al (2008) and Cross et al. (2012). 

The Pan-STARRS1 Surveys (PS1) and the PS1 public science archive have
been made possible through contributions by the Institute for
Astronomy, the University of Hawaii, the Pan-STARRS Project Office,
the Max-Planck Society and its participating institutes, the Max
Planck Institute for Astronomy, Heidelberg and the Max Planck
Institute for Extraterrestrial Physics, Garching, The Johns Hopkins
University, Durham University, the University of Edinburgh, the
Queen's University Belfast, the Harvard-Smithsonian Center for
Astrophysics, the Las Cumbres Observatory Global Telescope Network
Incorporated, the National Central University of Taiwan, the Space
Telescope Science Institute, the National Aeronautics and Space
Administration under Grant No. NNX08AR22G issued through the Planetary
Science Division of the NASA Science Mission Directorate, the National
Science Foundation Grant No. AST-1238877, the University of Maryland,
Eotvos Lorand University (ELTE), the Los Alamos National Laboratory,
and the Gordon and Betty Moore Foundation.

This project used data obtained with the Dark Energy Camera (DECam)
and the NOAO Data Lab, The Data Lab is operated by the National
Optical Astronomy Observatory, the national center for ground-based
nighttime astronomy in the United States operated by the Association
of Universities for Research in Astronomy (AURA) under cooperative
agreement with the National Science Foundation.

This publication makes use of data products from the Wide-field
Infrared Survey Explorer, which is a joint project of the University
of California, Los Angeles, and the Jet Propulsion
Laboratory/California Institute of Technology, and NEOWISE, which is a
project of the Jet Propulsion Laboratory/California Institute of
Technology. WISE and NEOWISE are funded by the National Aeronautics
and Space Administration.

CasJobs was originally developed by the Johns Hopkins University/
Sloan Digital Sky Survey (JHU/SDSS) team. With their permission, MAST
used version 3.5.16 to construct CasJobs-based tools for GALEX,
Kepler, the Hubble Source Catalog, and PanSTARRS.

This research has made use of the SVO Filter Profile Service
(http://svo2.cab.inta-csic.es/theory/fps/) supported from the Spanish
MINECO through grant AyA2014-55216 
The SVO Filter Profile Service\footnote{Rodrigo, C., Solano, E., Bayo, A. http://ivoa.net/documents/Notes/SVOFPS/index.html}
describes the Spanish VO Filter Profile Service. 
The Filter Profile Service Access Protocol. Rodrigo, C., Solano, E. http://ivoa.net/documents/Notes/SVOFPSDAL/index.html


\appendix

\section{Near-Infrared WFCAM Science Archive SQL queries}\label{sec:SQL_WSA}
Here we give the recipe and SQL that returned the near-infrared photometry 
for the VH$z$Qs from the  WFCAM Science Archive. 

The data are on the WFCAM Science Archive: \href{wsa.roe.ac.uk}{\tt wsa.roe.ac.uk}. 
Access the User Login form \href{WFCAM Science Archive}{\tt wsa.roe.ac.uk/login.html} 
with these credentials::
\begin{itemize}
    \item Username: {\tt WSERV1000} 
    \item password: {\tt highzqso} 
    \item community: {\tt nonsurvey}
\end{itemize}
Then going to the
\href{http://wsa.roe.ac.uk:8080/wsa/SQL_form.jsp}{{\tt Free Form SQL
Query}} page the Database release {\tt WSERV1000v20200113} can be
accessed which contains all the data we use here.

We {\it nota bene} a few things. First, the quantity {\tt aperJky3}
and {\tt aperJky3Err} are found in the {\tt wserv1000MapRemeasAver}
and {\tt wserv1000MapRemeasurement}, so care has to be taken to return
unique column names (otherwise e.g.
\href{http://docs.astropy.org/en/stable/io/fits/}{astropy.io.fits}
will crash).  As such, we alias {\tt aver.aperJky3} to {\tt
aperJky3Aver} and likewise for the error quantity. Aliases will be
necessary in some cases anyway, because some queries can be done
sensibly on multiple instances of the same table. Other times, one may
join tables on quantities such as {\tt catalogueID} or {\tt
apertureID}, where you are meaning the same thing, but aliases are then necessary 
for SQLServer to correctly comprehend the query. 

Second, the {\tt RA} and {\tt DEC} values returned by the WSA are in radians, if
used directly. To return values in degrees, use a selection with an alias, e.g. 
{\tt RA as RADeg} and {\tt DEC as DECDeg}. Other relevant query examples can be found in the
GitHub repository, under WSA\_VSA/SAMPLE\_SQL\_QUERIES, or the SQLCookbook in the WSA or VSA.

\onecolumn
\input{SQL_examples_WSA}
\twocolumn

\section{Near-Infrared VISTA Science Archive SQL queries}\label{sec:SQL_VSA}
In a very similar manner to the WSA, we give here the details on how to access
the VISTA Science Archive (VSA)

At the \href{http://horus.roe.ac.uk/vsa/login.html}{VSA Login}, enter 
with these credentials::
\begin{itemize}
    \item Username: {\tt VSERV1000} 
    \item password: {\tt highzqso} 
    \item community: {\tt proprietary}
\end{itemize}
Then the \href{http://horus.roe.ac.uk:8080/vdfs/VSQL_form.jsp}{Freeform SQL
Query} page is where the database release to use is {\tt
VSERV1000v20200113}. The queries are the same as those used in the
WSA, see Appendix~\ref{sec:SQL_WSA}, but the names of the tables starting
with {\tt wserv1000} should be replaced with {\tt vserv1000}.

\section{Updated VH$z$Q Sample}
\label{app:qso_sample}
Between the journal submitted version of this paper (which was
also the arXiv v1) we updated the sample of VH$z$Qs from a total of
463 objects to 488 objects.  16 of these objects were from
\cite{YangJ2018a} and 9 objects were added after discussions with
Krisztina Perger, with the relevant catalog found at 
\href{http://astro.elte.hu/$\sim$perger/catalog.html}{astro.elte.hu/$\sim$perger/catalog.html}.
We do not include the object DES0141-54 \citep{Belladitta2018}, since we were unable to 
obtain arcsecond level coordinates to perform our matched-aperture photometry.
We also update the positions of five objects, see Table~\ref{tab:disagree_objs}. 

One object in the Perger catalogue, VA-13492 we could not immediately cross-match. 
In this case the coordinates looked similar to SDSSJ1040-1155, and
the stated redshift was identical, $z=5.44$. Additionally the $i$-band magnitude 
was equivalent to SDSS J1040-1155
($i=25.9$ mag). SDSS J1040-1155 is first reported in \citet{Douglas2007} 
and a transcription error seems to have occurred in \citet{Flesch2015}.

\begin{table*}
\begin{tabular}{l r r r r r}
\hline  \hline
Object Name                      & R.A.               & Decl.  &  redshift  & difference \\
                                          & / deg             & / deg  &                & arcsec  \\
\hline
DELS J1048-0109               & +162.0795417 & --1.1611694        & 6.678   &   
12.1 \\
SHELLQs J1429-0002          & +217.3342500 & --0.0353889        & 6.04    & 
150.0 \\
DELS J1559+2212               & +239.7878750 & +22.2040083      & 6.07     &     3.0 \\
ATLAS J332.8017-32.1036 & +332.8017500 & --32.1035556      & 6.3394  &    3.2 \\
VIK J2318-3113                  & +349.5764625 & --31.2295417      & 6.4435  & 
239.9 \\
\hline  \hline
\end{tabular}
\caption{Five quasars with updated coordinates.}
\label{tab:disagree_objs}
\end{table*}


\bibliographystyle{mnras}
\bibliography{tester_mnras}

\end{document}

%% file: format.tex
\usepackage{amsmath,, amssymb}
\usepackage{bm}
\usepackage{blindtext}
\usepackage{cancel, caption, color}
\usepackage{dcolumn}
\usepackage{epsfig, epsf}
\usepackage{fancyhdr}
\usepackage[bottom,flushmargin,hang,multiple,para]{footmisc}
\usepackage{graphicx}
\usepackage{lscape, longtable, listings}
\usepackage{multirow}
\usepackage{natbib}
\usepackage{pdflscape}
\usepackage{subfigure}
\usepackage{textcomp}
\usepackage{hyperref, ifthen}
\usepackage{verbatim}
\usepackage[usenames,dvipsnames]{xcolor}

\usepackage{etoolbox}
\makeatletter
\patchcmd\@combinedblfloats{\box\@outputbox}{\unvbox\@outputbox}{}{%
   \errmessage{\noexpand\@combinedblfloats could not be patched}%
}%
 \makeatother

\def\nat{Nat} \def\apjl{ApJ~Lett.} \def\apj{ApJ}
\def\apjs{ApJS} \def\aj{AJ} \def\mnras{MNRAS}

\def\pasp{PASP} \def\aap{Astron.~\&~Astrophys.} \def\araa{ARA\&A}
\def\pasa{PASA}
\def\jcap{\ref@jnl{J. Cosmology Astropart. Phys.}}%

\def \gsim { \lower .75ex \hbox{$\sim$} \llap{\raise .27ex \hbox{$>$}} }
\def \lsim { \lower .75ex \hbox{$\sim$} \llap{\raise .27ex \hbox{$<$}} }

\newcommand{\cii}{C\,{\sc ii}\ }

\newcommand{\civ}{C\,{\sc iv}\ }

\newcommand{\mgii}{Mg\,{\sc ii}\ }

\newcommand{\nv}{N\,{\sc v}\ }

%% file: output_table_top5pc.tex
\begin{table}
\begin{tabular}{llrrc cccc cccc}
 \hline
 \hline
  \multirow{2}{*}{Survey} &  \multirow{2}{*}{QsoName} &   R.A. / deg  &   Decl. / deg  &  \multirow{2}{*}{redshift}   &  \multirow{2}{*}{Y}  &  \multirow{2}{*}{J}   &  \multirow{2}{*}{H}  &  \multirow{2}{*}{K}     &  \multicolumn{2}{c}{unWISE}  &  \multicolumn{2}{c}{AllWISE} \\ 
                          &                           &   (J2000)     &  (J2000)       &                              &                      &                       &                      &                         &          W1       & W2       & W3   & W4 \\ 
  \hline
  \hline
  \\
PSO & J000.3401+26.8358 & 0.34011 & 26.83588 & 5.75 & --- & $19.32\pm0.048$ &
--- & ---  &   $18.949\pm0.026$   &  $18.80\pm0.050$   &   $17.74\pm0.49$   & $>15.42$   \\
SDSS & J0002+2550 & 0.66412 & 25.84304 & 5.82 & --- & $19.37\pm0.087$ & --- &
--- &   $18.919\pm0.026$   &  $18.70\pm0.047$   &   $17.56\pm0.42$   &   $>15.34$   \\ 
SDSS & J0005-0006 & 1.46808 & --0.1155 & 5.85 & $20.70\pm0.211$ & $20.73\pm0.177$
& $20.05\pm0.082$ & $20.49\pm0.140$ &   $20.162\pm0.079$   &  $19.98\pm0.153$  
& $>17.59$   &   $>15.67$   \\ 
PSO & J002.1073-06.4345 & 2.10739 & --6.43456 & 5.93 & --- & $20.26\pm0.040$ &
--- & $19.74\pm0.082$ &   $19.471\pm0.044$   &  $19.24\pm0.078$   &   $>17.04$  
&   $>15.42$   \\
SDWISE & J0008+3616 & 2.21429 & 36.27041 & 5.17 & --- & $19.33\pm0.079$ & --- &
--- &   $18.687\pm0.021$   &  $18.71\pm0.044$   &   $>17.19$   &   $>15.45$   \\
PSO & J002.3786+32.8702 & 2.3787 & 32.87026 & 6.1 & --- & $22.17\pm0.855$ & ---
& --- &   $20.620\pm0.106$   &  ---  &   ---   &   ---   \\ 
SDSS & J0012+3632 & 3.137 & 36.53781 & 5.44 & --- & $19.01\pm0.061$ & --- & ---
& $18.490\pm0.017$   &  $18.51\pm0.036$   &   $17.15\pm0.23$   &   $15.35\pm0.3$   \\ 
SDSS & J0017-1000 & 4.31117 & --10.0154 & 5.01 & $19.24\pm0.032$ &
$19.26\pm0.045$ & --- & $18.78\pm0.065$ &   $18.645\pm0.022$   & 
$18.60\pm0.045$ & $17.17\pm0.33$   &   $>15.18$   \\ 
PSO & J004.3936+17.0862 & 4.39361 & 17.0863 & 5.8 & --- & $20.74\pm0.075$ & ---
& $20.28\pm0.114$ &   $20.503\pm0.103$   &  $19.98\pm0.145$   &  ---  &   --- \\ 
PSO & J004.8140-24.2991 & 4.81408 & --24.29916 & 5.68 & $19.54\pm0.048$ &
$19.19\pm0.049$ & --- & $19.07\pm0.090$ &   $18.989\pm0.028$   & 
$18.96\pm0.061$ &   $17.27\pm0.34$   &   $>15.48$   \\ 
VDES & J0020-3653 & 5.13112 & --36.89495 & 6.9 & --- & $20.42\pm0.100$ & --- &
$19.32\pm0.096$ &   $19.536\pm0.041$   &  $19.69\pm0.102$   &   $>17.83$   &  
$>15.00$   \\
SDSS & J0023-0018 & 5.87779 & --0.31011 & 5.06 & $20.62\pm0.148$ &
$20.50\pm0.083$ & $20.25\pm0.118$ & $20.27\pm0.102$ &   $19.518\pm0.046$   & 
$19.41\pm0.092$   & $>17.02$   &   $>15.15$   \\ 
PSO & J006.1240+39.2219 & 6.12404 & 39.22193 & 6.62 & --- & $21.28\pm0.483$ &
--- & --- &   $20.033\pm0.064$   &  ---   &   ---  &   --- \\
SDWISE & J0025-0145 & 6.36183 & --1.75903 & 5.07 & $18.03\pm0.020$ &
$17.95\pm0.020$ & $17.74\pm0.004$ & $17.59\pm0.022$ &   $17.520\pm0.009$   & 
$17.51\pm0.018$   &   $16.54\pm0.22$   &   $>15.17$   \\
PSO & J007.0273+04.9571 & 7.02733 & 4.95712 & 6.0 & $20.22\pm0.088$ &
$20.04\pm0.108$ & $20.40\pm0.196$ & $20.27\pm0.188$ &   $19.847\pm0.060$   & 
$19.89\pm0.135$   &   $>17.40$   &   $>14.98$   \\
SDWISE & J0031+0710 & 7.85775 & 7.17692 & 5.33 & $20.03\pm0.099$ &
$20.20\pm0.206$ & $19.49\pm0.141$ & $19.61\pm0.157$ &   $19.327\pm0.039$   & 
$18.96\pm0.063$   &   $>17.34$   &   $>15.06$   \\
CFHQS & J0033-0125 & 8.2975 & --1.42358 & 6.13 & --- & $21.41\pm0.264$ &
$21.32\pm0.335$ & $20.79\pm0.206$ &   $20.874\pm0.142$   &  ---   &   ---    &  
---   \\
SDSS & J0034+3759 & 8.55979 & 37.99833 & 5.63 & --- & $19.70\pm0.113$ & --- &
--- &   $19.149\pm0.029$   &  $19.01\pm0.056$   &   $>17.18$   &   $>15.76$   \\
PSO & J009.3573-08.1190 & 9.35733 & --8.11902 & 5.72 & $20.13\pm0.050$ &
$19.88\pm0.063$ & --- & $19.80\pm0.138$ &   $19.489\pm0.042$   & 
$19.48\pm0.095$ &   $>17.34$   &   $>15.02$   \\
DELS & J0038-1527 & 9.65042 & --15.45656 & 7.02 & --- & --- & --- & --- &  
$19.410\pm0.041$   &  $19.59\pm0.104$   &   $>17.45$   &   $>15.36$   \\
PSO & J009.7355-10.4316 & 9.73551 & --10.43164 & 6.0 & --- & --- & --- & --- &  
$19.160\pm0.032$   &  $19.00\pm0.061$   &   $>17.35$   &   $>15.07$   \\
PSO & J011.3899+09.0325 & 11.38987 & 9.03249 & 6.42 & $21.04\pm0.304$ & --- &
$20.64\pm0.244$ & $20.76\pm0.336$ &   $20.442\pm0.104$   &  ---  &   $>17.13$  
& $>15.44$   \\
PSO & J0046-2837 & 11.59854 & --28.62982 & 5.99 & $21.25\pm0.125$ &
$20.89\pm0.079$ & $20.61\pm0.112$ & $20.22\pm0.111$ & $19.732\pm0.050$ & $19.68\pm0.104$ & $>17.46$ & $>15.35$ \\
CFHQS & J0050+3445 & 12.52777 & 34.75601 & 6.25
& --- & $19.97\pm0.151$ & --- & ---  &   $19.250\pm0.033$   &  $18.99\pm0.055$  
&   $>18.08$   &   $>15.96$
\\
    \hline
    \hline
    \end{tabular}
    \caption{The first 24 (i.e. 5\%) of 488 very high-$z$ quasars in Right
    Ascension order with near and mid-infrared photometry.
                  The full table can be found \href{https://github.com/d80b2t/VHzQ/tree/master/data}{here}.
                  The AB magnitude system is used with the WISE Vega to AB offsets being ($\Delta$W1, $\Delta$W2, $\Delta$W3 $\Delta$W4)=(2.669, 3.281, 5.148, 6.66)
                  Since none of the first 24 objects have $Z$-band detections,
                  we don't report that column here (but is reported in the main table).
                  WISE AllWISE W3 and W4 values without formal errors are
                  low-SNR detections. While the number of significant figure
                  shown is larger than absolutely necesssary, it is useful to
                  verify the WFCAM to VISTA global offsets, see
                  Table~\ref{tab:WFCAM_vs_VISTA}.}
     \label{tab:output_table}
     \end{table}

%% file: SQL_examples_WSA.tex
\lstset{upquote=true}

\noindent
{\bf The following SQL will return 
the best magnitude in each band for each QSO.}

\begin{lstlisting}[
           language=SQL,
           showspaces=false,
           basicstyle=\ttfamily,
           numbers=left,
           numberstyle=\tiny,
           commentstyle=\color{gray}
        ]
SELECT 
qso.qsoName,  qso.ra as raJ2000, qso.dec as decJ2000, 
aver.apertureID,  aver.aperJky3 as aperJky3Aver, 
aver.aperJky3Err as aperJky3AverErr, aver.sumWeight, 
aver.ppErrBits as ppErrBitsAver, m.mjdObs, 
m.filterID, remeas.aperJky3, 
remeas.aperJky3Err, 
w.weight, remeas.ppErrBits, 
m.project

FROM 
finalQsoCatalogue as qso,  
MapApertureIDshighzQsoMap as ma,  
wserv1000MapRemeasAver as aver,  
wserv1000MapRemeasurement as remeas,  
MapProvenance as v,  
wserv1000MapAverageWeights as w, 
MapFrameStatus as mfs, 
Multiframe as m  

WHERE 
qso.qsoID=ma.objectID and 
ma.apertureID=aver.apertureID and 
aver.apertureID=remeas.apertureID and 
aver.catalogueID=v.combicatID and 
v.avSetupID=0 and 
v.catalogueID=remeas.catalogueID and 
w.combicatID=v.combicatID and 
w.catalogueID=v.catalogueID and 
w.apertureID=aver.apertureID and 
mfs.catalogueID=remeas.catalogueID and 
m.multiframeID=mfs.multiframeID and 
mfs.programmeID=10999 and 
mfs.mapID=1 
order by v.combicatID, m.mjdObs
\end{lstlisting}